\documentclass[
aps,
prb,
reprint,
showkeys,
twocolumn,
nofootinbib,
superscriptaddress,
]{revtex4-2}

\usepackage{natbib}
\usepackage{amssymb}
\usepackage{color}
\usepackage{amsfonts}
\usepackage{textcomp}
\usepackage{amsmath}
\usepackage[charter]{mathdesign}
\usepackage{easyReview}
\usepackage{microtype} 
\usepackage{upgreek}
\usepackage{booktabs}
\newcommand{\ped}[1]{\ensuremath{_{\rm #1}}}
\newcommand{\apex}[1]{\ensuremath{^{\rm #1}}}
\definecolor{link}{RGB}{57,106,177}
\definecolor{link}{RGB}{0,0,0}
\definecolor{darkgreen}{RGB}{0,128,0}
\usepackage{float}
\usepackage{graphicx}
\usepackage{hyperref}
\hypersetup{
	colorlinks=true,
	linkcolor=link,
	citecolor=link,
	urlcolor=link
}

\begin{document}

\title{Direct evidence for two-gap superconductivity in hydrogen-intercalated titanium diselenide}
	
\author{Erik Piatti}
\email{erik.piatti@polito.it}
\affiliation{Department of Applied Science and Technology, Politecnico di Torino, I-10129 Torino, Italy}
\author{Gaia Gavello}
\affiliation{Department of Applied Science and Technology, Politecnico di Torino, I-10129 Torino, Italy}
\author{Giovanni A. Ummarino}
\affiliation{Department of Applied Science and Technology, Politecnico di Torino, I-10129 Torino, Italy}
\author{Filip Ko{\v s}uth}
\affiliation{Centre of Low Temperature Physics, Institute of Experimental Physics, Slovak Academy of Sciences, SK-04001 Ko{\v s}ice, Slovakia}
\affiliation{Centre of Low Temperature Physics, Faculty of Science, P. J. {\v S}af{\'a}rik University, SK-04001 Ko{\v s}ice, Slovakia}
\author{Pavol Szab{\'o}}
\author{Peter Samuely}
\affiliation{Centre of Low Temperature Physics, Institute of Experimental Physics, Slovak Academy of Sciences, SK-04001 Ko{\v s}ice, Slovakia}
\author{Renato S. Gonnelli}
\email{renato.gonnelli@polito.it}
\affiliation{Department of Applied Science and Technology, Politecnico di Torino, I-10129 Torino, Italy}
\author{Dario Daghero}
\affiliation{Department of Applied Science and Technology, Politecnico di Torino, I-10129 Torino, Italy}


\begin{abstract}
Transition-metal dichalcogenides (TMDs) offer an extremely rich material platform in the exploration of unconventional superconductivity.
The unconventional aspects include exotic coupling mechanisms such as the Ising pairing, a complex interplay with other electronic orders such as charge-density waves (CDWs), symmetry-breaking and topological effects, and non-trivial gap structures such as multi-gap and possible nodal phases.
Among TMDs, titanium diselenide (1$T$-TiSe$_2$) is one of the most studied and debated cases.
Hints to an anomalous structure of its superconducting order parameter have emerged over the years, possibly linked to its spatial texturing in real and reciprocal space due to the presence of a 2$\times$2$\times$2 CDW phase, or to a pressure-driven multi-band Fermi surface.
However, a direct evidence for a non-trivial structure of the superconducting gap in this material is still lacking.
In this work, we bring the first evidence for a two-gap structure in the recently-discovered H-intercalated TiSe$_2$ superconductor (with T$_c \simeq 3.6$ K) by an extensive experimental study that combines magnetotransport measurements, point-contact spectroscopy and scanning tunnel spectroscopy.
We show that the temperature dependence of the upper critical field (for $\vec B \parallel c$) as well as the shape of the point-contact and tunneling spectra strongly suggest the existence of two distinct superconducting gaps, and can indeed all be fitted in a self-consistent way with the same gap amplitudes $\Delta_1 = 0.26 \pm 0.12$ meV and $\Delta_2 =0.62 \pm 0.18$ meV.
\end{abstract}

\maketitle
	

\noindent Transition metal dichalcogenides (TMDs) are a class of quasi-two-dimensional materials sharing the chemical fomula $MX_2$, where $M$ is a transition metal and $X$ is a chalcogen (S, Se, Te) atom, and exhibiting a layered crystal structure based on the stacking of covalently-bonded $MX_2$ trilayers which interact via weak inter-trilayer van der Waals forces\,\cite{Wilson1969AP}.
Similarly to cuprates\,\cite{Shen2008MT}, heavy-Fermion\,\cite{Stewart1984RMP} and iron-based\,\cite{Stewart2011RMP, Fernandes2022Nature} compounds, the vast chemical variety and complex electronic structures of TMDs often endows them with rich phase diagrams characterized by multiple quantum phases\,\cite{Manzeli2017NRM, Choi2017MT, Li2021NatCommun}, the most commonly-occurring ones being superconductivity and charge-density waves (CDW)\,\cite{Klemm2015PC}.
Additionally, the weak inter-layer bonding makes TMDs easily exfoliable down to the two-dimensional single-trilayer limit and eminently suited for chemical doping via intercalation of both atomic and molecular ions and neutral molecules\,\cite{Dresselhaus1987MRSB, Stark2019AdMa}.
This in turn provides ample control over their structural, electronic and optical properties, making TMDs remarkably appealing from both a fundamental and an application-oriented point of view\,\cite{Klemm2015PC, Wang2012NatNano, Geim2013Nature, Fiori2014NatNano, Voiry2015CSR, Piatti2021NatElectron}.\\
\indent Among TMDs, 1$T$-TiSe$_2$ has been one of the most extensively investigated, starting from the discovery of a commensurate 2$\times$2$\times$2 CDW phase below ${\approx}200$\,K in the undoped compound\,\cite{DiSalvo1976PRB, Rasch2008PRL} whose origin is still a matter of debate\,\cite{Rossnagel2002PRB, Kogar2017Science, Knowles2020PRL, Otto2021SciAdv, Lin2022PRL, Novko2022PRB}.
Interest in this compound was further invigorated by the discovery that Cu intercalation could suppress the commensurate CDW order in favor of the emergence of a superconducting (SC) dome, with critical temperatures up to $\approx 4.15$\,K\,\cite{Morosan2006NatPhys}.
A similar tunability was later reported also in the case of applied pressure\,\cite{Kusmartseva2009PRL}, electric-field-induced charge doping\,\cite{Li2016Nature}, and Pd\,\cite{Morosan2010PRB}, Li\,\cite{Liao2021NatCommun}, H\,\cite{Piatti2023CommunPhys}, and diamine\,\cite{Sato2017JPSJ} intercalation.
Significant attention was devoted in particular to the possibility that the spatial texture associated with the CDW phase could affect the anisotropy of the SC pairing or even result in a non-trivial symmetry of the SC gap, especially after the observation of the Little-Parks effect in electrostatically-gated and Li-doped TiSe\ped{2} indicated the existence of a periodic modulation in the amplitude of the SC order parameter\,\cite{Li2016Nature, Liao2021NatCommun}.\\
\indent Several works therefore aimed at probing the structure of the SC gap in TiSe$_2$. Most investigations have been carried out in the Cu-intercalated compound and concur in finding a single isotropic gap in the system\,\cite{Li2007PRL, Hillier2010PRB, Kacmarcik2013PRB}.
Some exceptions to this general consensus have however been reported, usually in TiSe$_2$ systems where CDW and SC orders were simultaneously present and the amplitude of the latter might therefore be spatially textured: These include a possible two-gap state detected by $\upmu$SR measurements in underdoped Cu$_x$TiSe\ped{2}\,\cite{Zaberchik2010PRB} and a zero-bias peak at odds with isotropic $s$-wave pairing reported in the point-contact spectra of electrostatically-doped TiSe\ped{2}\,\cite{Li2016Nature}.
Unconventional $s_\pm$ pairing mediated by CDW fluctuations and driven by a Lifshitz transition has also recently been proposed to be responsible for superconductivity in pressurized TiSe\ped{2}, and possibly extend to the electron-doped compounds as well\,\cite{Hinlopen2024SciAdv}.\\
\indent In this context, the recently-discovered H-intercalated TiSe\ped{2} superconductor\,\cite{Piatti2023CommunPhys} is a prime candidate to host a non-trivial structure of the SC order parameter.
This is because on the one hand the robust survival of CDW order detected at the intercalant sites by $^1$H nuclear magnetic resonance\,\cite{Prando2023PRMater} suggests a possible local coexistence of CDW and SC orders in the system; On the other hand, the large doping levels attained in H$_x$TiSe\ped{2} (up to $x{\approx}2$) enable a full reconstruction of the Fermi surface by an orbital-selective filling of the Ti $d_{z^2}$ band\,\cite{Piatti2023CommunPhys}, mimicking that of SC TMD compounds NbS\ped{2} and NbSe\ped{2} where highly-anisotropic distributions of the SC gap on the Fermi surface\,\cite{Heil2017PRL, Sanna2022npjQM} result in well known two-gap behaviors\,\cite{Guillamon2008PRL, Kacmarcik2010PC, Yokoya2001Science, Boaknin2003PRL, Guillamon2008PRB, Rahn2012PRB, Noat2015PRB}.\\
\indent In this work, we present clear evidence for two-gap superconductivity in H-intercalated TiSe\ped{2} revealed by directional point-contact Andreev-reflection spectroscopy (PCARS) and scanning tunneling spectroscopy (STS) measurements down to 0.3\,K, by the temperature-dependence of the upper critical magnetic field determined by magnetotransport measurements, and their combined quantitative analysis within the two-band BCS and Eliashberg theory.
We speculate that the two-gap structure may arise from an extremely anisotropic distribution of the SC gap function on the single Fermi surface of the material due to the orbital-selective filling induced by H intercalation.

\section*{Results}

\begin{figure*}
	\begin{center}
		\includegraphics[keepaspectratio, width=\textwidth]{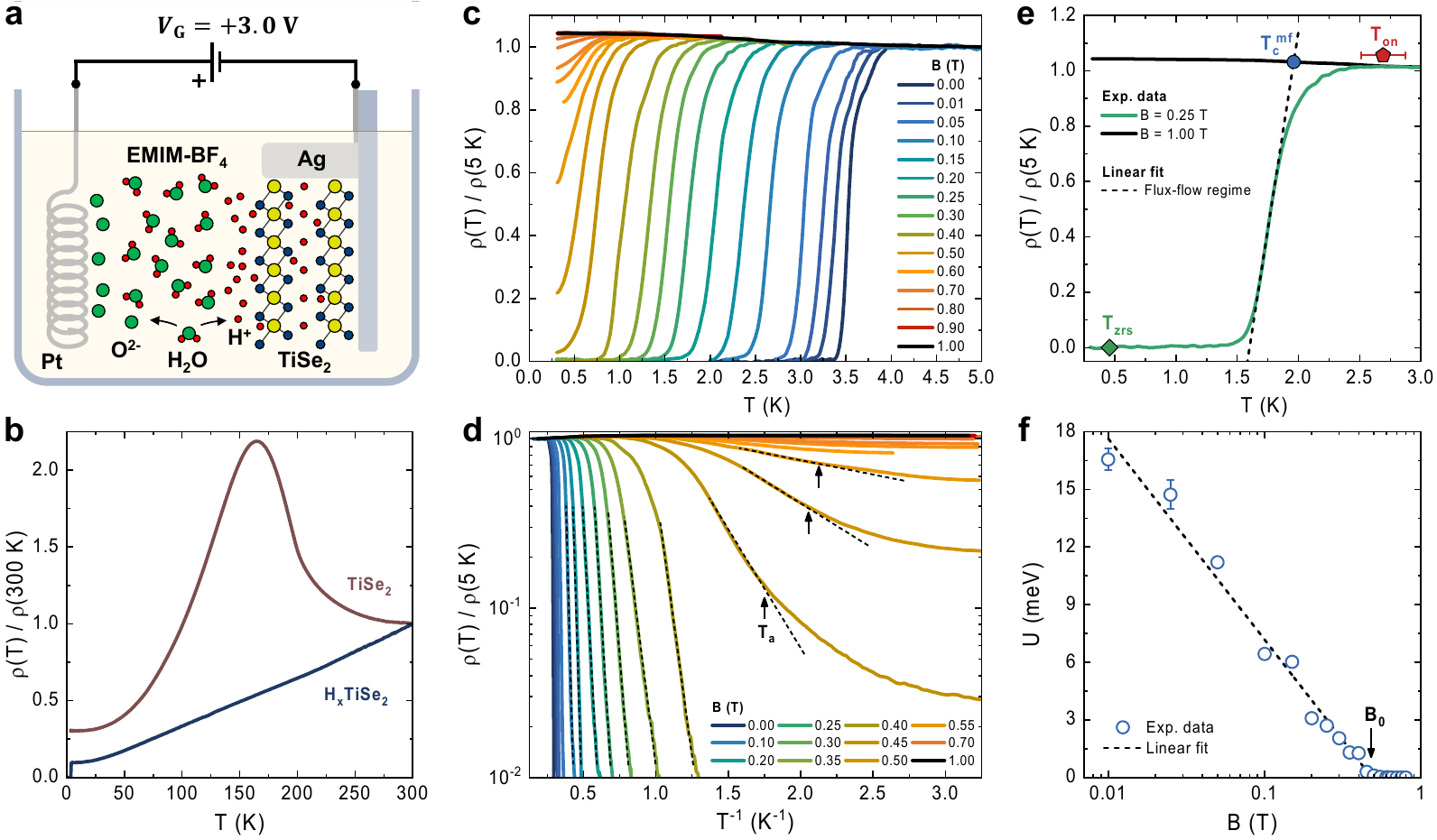}
	\end{center}\vspace{-15pt}
	\caption{
            Gate-driven H intercalation and superconductivity in TiSe\ped{2}.
            \textbf{a}, Schematic depiction of the experimental setup used in the gate-driven H intercalation process. The voltage $V\ped{G}$ applied between the Pt gate electrode and the TiSe\ped{2} crystal splits the water molecules absorbed by the EMIM-BF\ped{4} ionic liquid and drives the H\apex{+} ions inside the TiSe\ped{2} structure.
            \textbf{b}, Resistivity $\rho$ vs temperature $T$ in pristine and H-intercalated TiSe\ped{2} at zero magnetic field.
            \textbf{c}, $\rho(T)$ curves of H-intercalated TiSe\ped{2} across the SC transition for increasing perpendicular magnetic fields $B$.
            \textbf{d}, same as \textbf{c} plotted in semilogarithmic scale against $T^{-1}$. Dashed lines are linear fits to the data which highlight the TAFF scaling. Arrows indicate the temperatures $T\ped{a}$ where a crossover to pinning-free vortex motion occurs.
            \textbf{e},  $\rho(T)$ curve at $B=0.25$\,T with three characteristic $T$ scales explicitly indicated: The zero-resistance state temperature $T\ped{a}$, the mean-field critical temperature $T\ped{c}\apex{mf}$, and the onset temperature $T\ped{on}$. The error bar represents the uncertainty on $T\ped{on}$ due to the experimental resolution on the condition $d\rho/dT=0$. The dashed line is a linear fit to the data in the flux-flow regime. The normal-state $\rho(T)$ curve at $B=1$\,T is also reported.
            \textbf{f}, $B$ dependence of the activation barrier for vortex motion $U$ in semilogarithmic scale, obtained from the slope of the linear fits shown in \textbf{d}. The corresponding statistical uncertainties are reported as error bars. The dashed line is a linear fit to the data.
	}
	\label{fig:transport}
\end{figure*}

\noindent \textbf{Sample synthesis.}
Our magneto-transport and point-contact spectroscopy measurements were carried out on TiSe\ped{2} single crystals which were intercalated with H atoms via the ionic liquid gating method\,\cite{Piatti2023CommunPhys, Boeri2022JPCM}.
As illustrated in Fig.\,\ref{fig:transport}a, the intense electric field at the voltage-polarized electrolyte/TiSe\ped{2} interface dissociates the water molecules absorbed in the ionic liquid and injects the H\apex{+} ions in the van der Waals gap of the layered crystal\,\cite{Lu2017Nature, Meng2022PRB}, while the gating time is used as a knob to control the amount of non-volatile H incorporation (Methods).
Fig.\,\ref{fig:transport}b displays the resulting effects on the temperature ($T$) dependence of the in-plane resistivity, $\rho(T)$.
Namely, H doping strongly suppresses the resistivity anomaly around ${\sim}160$\,K typical of undoped TiSe\ped{2} crystals (solid brown line), which occurs as CDW order partially gaps the Fermi surface\,\cite{Morosan2006NatPhys, Morosan2010PRB, Li2016Nature}; It also induces a more conventional metallic behavior in H$_x$TiSe\ped{2} (solid blue line) and the development of a SC transition below $4$\,K. These effects are accompanied by electron doping which removes the CDW-driven sign change in the Hall coefficient around ${\sim}215$\,K\,\cite{Piatti2023CommunPhys, DiSalvo1976PRB, Knowles2020PRL, Liao2021NatCommun}.
Despite the loss of these transport anomalies, residual CDW ordering persists in the H-doped crystals as evidenced by the strong $T$-dependence of the Hall density $n\ped{H}$\,\cite{Piatti2023CommunPhys} and the peculiar dynamics of the $^1$H nuclear magnetic moments\,\cite{Prando2023PRMater}.
Additional details can be found in Ref.\,\citenum{Piatti2023CommunPhys}, which established gate-driven protonation to be a reliable tool to control the ground state and induce SC in 1$T$-TiSe\ped{2}.\\

\noindent \textbf{Magneto-transport.}
We now focus on the response of the SC resistive transition of a H$_x$TiSe\ped{2} sample with $x\approx2$ and $n\ped{H}\approx5{\times}10^{20}$\,cm\apex{-3} (i.e., the maximum attainable H doping\,\cite{Piatti2023CommunPhys}) to the application of magnetic fields $B$ perpendicular to the $ab$ planes of the crystal.
Fig.\,\ref{fig:transport}c shows $\rho(T)$ normalized to its value at 5\,K, for zero and finite $B$ in linear scale. When $B=0$, the resistive transition reaches its midpoint -- i.e., the $T$ where $\rho(T)$ becomes half of its normal-state value -- at 3.53\,K, and drops below the noise level at $T\ped{zrs}=2.9\pm0.2$\,K, entering the zero-resistance state (ZRS).
On increasing $B$, the resistive transitions are shifted to lower $T$ and progressively broadened, likely owing to thermal phase fluctuations promoted by the inhomogeneous and quasi-2D character of the system\,\cite{Saito2018NatCommun, Li2019NanoLett}; nevertheless, a well-defined ZRS persists up to $B=0.25$\,T, whereas at higher fields dissipation is observed even at the lowest $T$ accessible in our setup (0.3\,K). The normal-state $\rho(T)$ curve, attained at $B=1$\,T, displays a slight upturn, likely caused by the electronic correlations\,\cite{Altshuler1980, FukuyamaBook} arising from the localized Ti $d$ orbitals\,\cite{Bianco2015PRB, Rohwer2011}.
As highlighted by the dashed lines in Fig.\,\ref{fig:transport}d, which plots $\rho(T)/\rho(\mathrm{5\,K})$ in semilogarithmic scale as a function of $T^{-1}$, the dissipative regime can be well-described in terms of thermally-activated flux flow (TAFF) of vortices, $\rho(H,T)=\rho_0 \exp{\left[-U(B)/k\ped{B}T\right]}$, where $U(B)$ is the potential barrier to vortex motion created by the pinning centers\,\cite{Blatter1994RMP}.
With respect to our earlier findings in the same system\,\cite{Piatti2023CommunPhys}, however, the lower temperatures investigated here allow us to observe a range of $T$ and $B$ where $\rho(T)$ remains larger than predicted by the TAFF model, and we define $T\ped{a}(B)$ as the onset of this departure.\\
\indent An accurate determination of the $B-T$ phase diagram of layered superconductors also requires an estimation of the $T$-dependence of the magnetic field of SC onset, $B\ped{on}(T)$, and of the mean-field upper critical field, $B\ped{c2}\apex{mf}(T)$.
The $B\ped{on}(T)$ curve can be constructed by determining the crossing points between the $\rho(B)$ isotherms obtained by matrix inversion of the $\rho(T)$ curves\,\cite{Lewellyn2019PRB, Wang2023PRB} (see Supplementary Note\,1) or, equivalently, by plotting the field dependence of the onset critical temperature, $T\ped{on}(B)$, defined as the value of $T$ at which, for any given field $B$, $d\rho/dT=0$\,\cite{Saito2018NatCommun}.     
The $B\ped{c2}\apex{mf}(T)$ curve is especially necessary for a proper understanding of the vortex state and for the assessment of the gap symmetry.
In 2D and highly-anisotropic superconductors, the SC transition 
is often smeared by thermal fluctuations and inhomogeneities, which in turn make commonly-used criteria (such as taking the point where the resistivity is 90\% of the normal-state one) unreliable\,\cite{Saito2018NatCommun, Theunissen1997PRB}. 
Therefore, we adopted the approach by Berghuis and Kes\,\cite{Berghuis1993PRB} and linearly extrapolated the resistivity in the flux-flow regime to the normal state.
Doing so on the $\rho(B)$ isotherms (as exemplified in Supplementary Note\,1) directly gives the $B\ped{c2}\apex{mf}(T)$ curve, while using the $\rho(T)$ curves at fixed $B$ (as shown Fig.\,\ref{fig:transport}e for $B=0.25$\,T) provides the $T\ped{c}\apex{mf}(B)$ curve.\\
\begin{figure}
	\begin{center}
		\includegraphics[keepaspectratio, width=\columnwidth]{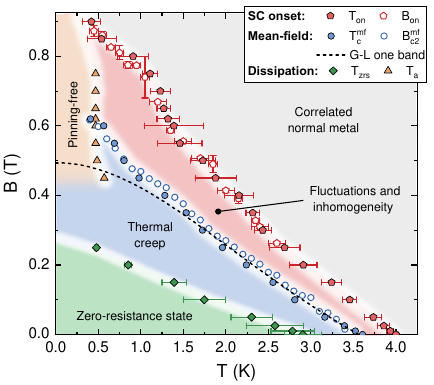}
	\end{center}\vspace{-15pt}
	\caption{
		Magnetic field-temperature ($B{-}T$) phase diagram of H-intercalated TiSe\ped{2} determined from the magneto-transport measurements.
        $T\ped{zrs}$ (green diamonds) are the temperatures below which zero resistance is observed.
        $T\ped{a}$ (orange triangles) mark the boundary for the dissipative state due to pinning-free vortex motion.
        $T\ped{on}$ and $B\ped{on}$ (filled and open red pentagons) show the SC onset as defined by $d\rho/dT=0$ and the crossing points of the $\rho(B)$ isotherms, respectively.
        $T\ped{c}\apex{mf}$ and $B\ped{c2}\apex{mf}$ (filled and open blue circles) represent the mean-field upper critical field determined by linear extrapolation to the normal state of the $\rho(T)$ and $\rho(B)$ curves, respectively\,\cite{Berghuis1993PRB}.
        Error bars are the resolution on the definitions due to the experimental noise level.
        The black dashed line shows the expected mean-field line according to the one-band Ginzburg-Landau model\,\cite{Tinkham1963PR}.
        The truly dissipationless SC state is observed for $T \leq T\ped{zrs}$ (green region).
        Dissipation develops due to thermal creep at moderate $B$ below the mean-field line (blue region) and free vortex motion at large $B$ where the pinning potential is suppressed at low $T$ (orange region).
        Inhomogeneity and amplitude fluctuations dominate between the mean-field line and the SC onset (red region).
	}
	\label{fig:phase_diagram}
\end{figure}
\indent Fig.\,\ref{fig:phase_diagram} displays the resulting comprehensive $B{-}T$ phase diagram, which comprises different regions bounded by the zero-resistance curve ($T\ped{zrs}(B)$, green diamonds), the mean-field critical curve ($T\ped{c}\apex{mf}(B)$ and $B\ped{c2}\apex{mf}(T)$, filled and hollow blue circles), and the two crossover curves associated with the SC onset ($T\ped{on}(B)$ and $B\ped{on}(T)$, filled and hollow red pentagons) and the departure from TAFF vortex motion ($T\ped{a}(B)$, orange triangles) respectively.
Both $T\ped{on}(B)$ and $B\ped{on}(T)$ lie well above the mean-field curve, indicating that a large region exists between the onset of SC fluctuations and the actual mean-field transition -- as commonly observed in quasi-2D superconductors.
The interval between the onset field and the mean-field one increases as $T$ is reduced, which is not expected for pure thermal fluctuations of the order parameter amplitude\,\cite{Theunissen1997PRB, Ullah1990PRL} and indicates an important role played by an underlying mesoscopic inhomogeneity\,\cite{Li2019NanoLett, Dezi2018PRB, Venditti2019PRB}, possibly assisted by quantum fluctuations as $B$ is increased\,\cite{Saito2018NatCommun}.
The region between the mean-field line and the ZRS line is instead dominated by collective flux creep (thermal phase fluctuations)\,\cite{Blatter1994RMP}: This is evidenced by the fact that the $B$ dependence of the activation barrier for vortex motion scales as $U(B)=U_0\ln{(B_0/B)}$ (as shown in Fig.\,\ref{fig:transport}f) with $U_0 = 53\pm2$\,meV and $B_0 = 0.48\pm0.06$\,T, comparable to earlier reports in both ion-gated\,\cite{Li2019NanoLett} and H-doped\,\cite{Piatti2023CommunPhys} TiSe\ped{2}.
Thermal creep therefore breaks down when $U(B)\rightarrow 0$ for $B\gtrsim B_0$, which indeed corresponds to the pocket below the $T\ped{a}(B)$ curve: This implies that the pinning potential effectively vanishes at high $B$, leading to a pinning-free motion of vortices which we ascribe to a combination of finite excitation current, high-frequency noise, and quantum fluctuations\,\cite{Xing2021NanoLett}.\\
\begin{figure*}
	\begin{center}
		\includegraphics[keepaspectratio, width=\textwidth]{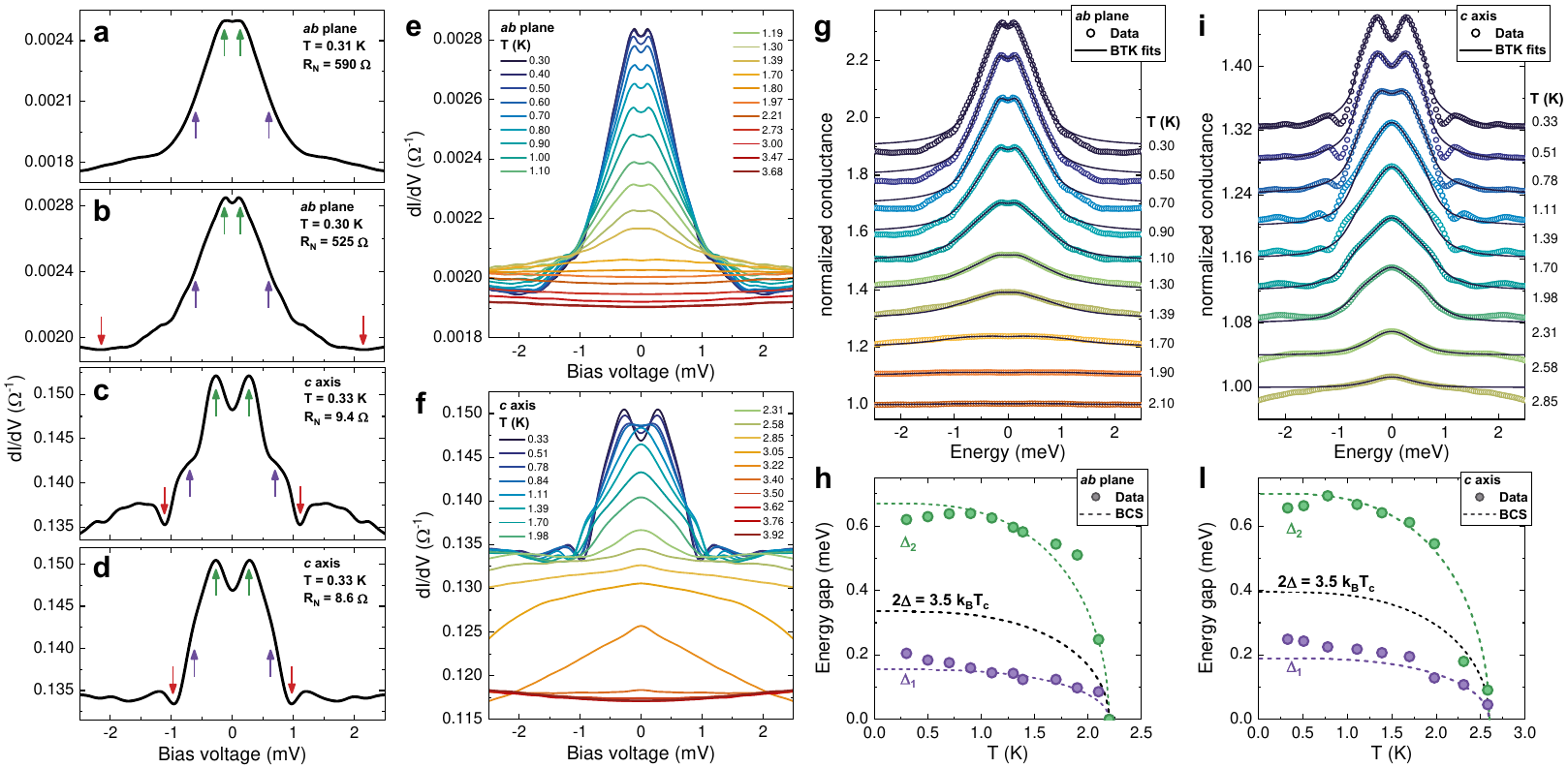}
	\end{center}\vspace{-15pt}
	\caption{
           Directional PCARS measurements in H$_x$TiSe\ped{2}.
            \textbf{a-d}, Examples of symmetrized differential conductance curves ($dI/dV$ vs. $V$) measured in $ab$-plane (\textbf{a,b}) and $c$-axis (\textbf{c,d}) contacts. Normal-state resistance $R\ped{N}$ and temperature $T$ are indicated in the labels. Green, violed and red arrows highlight the spectral features associated to the small gap $\Delta_1$, large gap $\Delta_2$, and critical current "dips".
            \textbf{e,f}, $T$ dependence of the $dI/dV$ spectra displayed in panels \textbf{b,d}.
            \textbf{g}, Same curves of panel \textbf{e} after normalization (symbols) and the best-fitting curves (line) according to the two-band BTK model.
            \textbf{h}, $T$ dependence of $\Delta_1$ (purple circles) and $\Delta_2$ (green circles) extracted from the fits in panel \textbf{g}. Dashed lines are functions of the form $\Delta_i(T) = \Delta_i(0)\tanh \left(1.74 \sqrt{ T\ped{c}\apex{A}/T-1}\right)$.
            \textbf{i,l}, Same as \textbf{g,h} for the curves of panel \textbf{f}.
	}
	\label{fig:PCARS}
\end{figure*}
\indent Finally, we focus on the mean-field critical curve. As already observed\,\cite{Saito2018NatCommun}, the mean-field critical temperature $T\ped{c}\apex{mf}$ is close to the midpoint of the resistive transition in $\rho(B)$ when $B\rightarrow 0$ but, on increasing $B$, it approaches the value of $T$ where $\rho$ is 95\% of the normal-state value.
Moreover, the mean-field line retains an upward curvature in the whole $T$ range, exactly as the onset curve.
This behavior is at odds with what observed in gated ZrNCl\,\cite{Saito2018NatCommun} where the SC onset curve displayed a low-$T$ upturn, but the mean-field curve exhibited a clear downward curvature, compatible with the Wertharmer–Helfand–Hohenberg (WHH) theory. 
Incidentally, this means that, in our system, the upward curvature of the SC onset line is not (only) due to the effects of inhomogeneities and fluctuations as in ZrNCl, but reflects an intrinsic property of the material which already shows up in the $T$ dependence of $B\ped{c2}\apex{mf}$.
For $T< T\ped{c}/2$ such a $T$ dependence is clearly incompatible with both the WHH theory and the standard one-band Ginzburg-Landau dependence $B\ped{c2}\propto(1-t^2)/(1+t^2)$, where $t=T/T\ped{c}$\,\cite{Tinkham1963PR} (black dashed line in Fig.\,\ref{fig:phase_diagram}).
Conversely, positive curvatures in $B\ped{c2}(T)$ are usually a hallmark of superconductors where at least two effective SC gaps open on the Fermi surface\,\cite{Gurevich2003PRB, Xing2017SciRep, Ding2022NL}.\\

\noindent \textbf{Point-contact Andreev-reflection spectroscopy.}
We directly investigated the SC gap structure in our H$_x$TiSe$_2$ crystals by means of directional PCARS measurements\,\cite{Naidyuk2005Book, Daghero2010SUST} that we performed by means of the so-called soft technique (Methods).
The direction of (main) current injection was controlled by making the point contacts either on the top surface (\textit{c}-axis contacts) or the side (\textit{ab}-plane contacts) of the platelet-like crystals.
In both cases, we obtained conductance curves with spectroscopic signals only when the contacts where made immediately after cleaving (\textit{c} axis) or breaking/cutting (\textit{ab} plane) the crystals to minimize the exposure of the probed surfaces to air and moisture.\\
\indent Fig.\,\ref{fig:PCARS} shows two representative examples of differential conductance spectra ($dI/dV$ vs $V$) of $ab$-plane point contacts (panels a,b) and $c$-axis point contacts (panels c,d).
All the spectra were acquired at about 0.3\,K and in zero magnetic field.
The resistance of the contacts that allowed the detection of good spectroscopic signals is significantly different for the two directions, being much larger for $ab$-plane contacts ($R\ped{N} \gtrsim 500 \Omega$) than for $c$-axis ones ($R\ped{N} \simeq 10 \Omega$).
All the spectra display zero-bias dips that are suggestive of nodeless SC gaps; in no cases we obtained the zero-bias peaks or cusps that are typical of nodal gaps or even gaps with zeros\,\cite{Daghero2010SUST, Torsello2022npjQM}.
The two low-energy conductance maxima (indicated by green arrows) can definitely be associated to a SC gap, but these are not the only feature of the curves.
Indeed, they all show additional structures, in the form of smooth shoulders or kinks (indicated by purple arrows) which can be interpreted as the hallmark of another gap.
Similar structures were indeed often found in  multigap superconductors like MgB$_2$\,\cite{Gonnelli2002PRL, Gonnelli2004PRB} and iron-based compounds\,\cite{Daghero2011RoPP, Daghero2020PRB, Torsello2022npjQM}.
It can be clearly seen that these features occur approximately at the same energy irrespective of the direction of current injection and of the contact resistance, which indicates their intrinsic nature.\\
\indent Additional structures, indicated by red arrows, can also be identified.
In the case of $c$-axis contacts, the shape of these structures allows identifying them with the typical "dips" that arise when the contact is not in the purely ballistic regime and are observed when the current density in the contact becomes overcritical, locally driving the superconductor in the resistive state\,\cite{Daghero2023LTP, Daghero2010SUST, Sheet2004}.
In $ab$-plane contacts these dips are much weaker and occur at higher bias voltages, as expected due to the much larger resistance of the contacts; in Fig.\,\ref{fig:PCARS}a they even fall outside the voltage range.\\
\indent Fig.\,\ref{fig:PCARS}e and f show the $T$ dependence of the conductance curves in Fig.\,\ref{fig:PCARS}b and d.
In either case, increasing $T$ makes the amplitude of the Andreev signal decrease, while the two conductance maxima progressively shift to lower bias and then merge in a single hump at zero bias, as expected for gap amplitudes that decrease upon heating.
The normal state is achieved when any trace of the low-energy conductance enhancement due to Andreev reflection disappears and subsequent curves collapse on a featureless spectrum.
In both cases, a $T$-dependent downward shift of the high-energy tails of the curves can also be seen. This is due to the onset of the so-called spreading resistance\,\cite{Chen2010PRB, Doring2014, Daghero2023LTP}, i.e. the resistance of the bulk of the crystal, that acts as an additional resistance in series to the contact in the proximity of the resistive transition measured by transport (see Fig.\,\ref{fig:transport}c), as already pointed out elsewhere\,\cite{Torsello2022npjQM, Piatti2024Nanomat}.
\\
\indent The fit of the curves in Fig.\,\ref{fig:PCARS}e and f is reported in Fig.\,\ref{fig:PCARS}g and i.
The raw conductance spectra were symmetrized and normalized, i.e. divided by the normal-state conductance curve measured just above $T\ped{c}$ and suitably corrected to get rid of the contribution of the spreading resistance\,\cite{Torsello2022npjQM, Piatti2024Nanomat}.
The resulting experimental spectra were then fitted by using a 2D version\,\cite{Kashiwaya2000RoPP, KashiwayaPRB1996} of the Blonder-Tinkham-Klapwijk (BTK) model\,\cite{BlonderPRB1982} further generalized to the case of two isotropic ($s$-wave) energy gaps\,\cite{Gonnelli2002PRL, Daghero2010SUST}.
The model includes the gap amplitudes $\Delta_i$, the broadening factors $\Gamma_i$, the interface barrier heights $Z_i$ and the spectral weights of the two gaps $w_i$ (bounded by $w_1 + w_2 = 1$) as free parameters, which can be univocally set thanks to the different effect that each parameter plays in the shape of the curve\,\cite{Daghero2010SUST, Daghero2011RoPP}.
We first tried to fit the curves with a single anisotropic gap, but we found that this model is unable to reproduce the shape of the spectra, especially in $c$-axis contacts (see Supplementary Note\,2).
As clearly seen in Fig.\,\ref{fig:PCARS}g and i, the two-gap model is instead able to grasp most of the features of the spectra, apart from those associated with the dips, which would require a more complicated model\,\cite{Daghero2023LTP} as shown in Supplementary Note\,3.
The amplitude of the gaps is actually unaffected by whether we include the dip or not in the fit, which is a good check of internal consistency.
The curves could be fitted only up to temperatures slightly above 2\,K, because (as seen in Fig.\,\ref{fig:PCARS}e,f) above this value of $T$ the spreading resistance starts to play a dominant role and the spectra, while being shifted downwards, are also stretched horizontally\,\cite{Chen2010PRB, Doring2014, Daghero2023LTP} preventing any spectroscopically-meaningful extraction of energy-resolved information.\\
\indent The $T$ dependence of the gap amplitudes, obtained through an automatic least-square procedure, are shown in Fig.\,\ref{fig:PCARS}h and l, whereas the $\Gamma$ and $Z$ parameters are summarized in Supplementary Note\,4
.
The smaller gap $\Delta_1$ decreases on increasing temperature, approximately following a BCS-like $\Delta(T)$ curve $\Delta_i(T) = \Delta_i(0)\tanh \left(1.74 \sqrt{ T\ped{c}\apex{A}/T-1}\right)$ (dashed purple line) -- even though, instead of saturating at low temperature, below approximately 0.7\,K it displays a linear behavior similar to that already observed in 1\textit{H}-NbSe$_2$ by STS\,\cite{zhao_2019NaturePhys} and in RbCa$_2$Fe$_4$As$_4$F$_2$ by soft PCARS\,\cite{Torsello2022npjQM}.
Surprisingly, the gap tends to zero well before the $T\ped{c}\apex{mf}$ measured by transport, i.e. at $T\ped{c}\apex{A}\approx2.2$\,K in Fig.\,\ref{fig:PCARS}h and 2.6\,K in Fig.\,\ref{fig:PCARS}l.
This latter feature is common to \emph{all} the contacts we made, irrespective of the direction of current injection, of the value of $R\ped{N}$ and of the region of the crystal where the contact was made.
Rather, these values are closer to the lower bound of the estimate for $T\ped{zrs}$, indicating a close link between $T\ped{c}\apex{A}$ and the onset of TAFF.
Notably, $T\approx2$\,K is also the temperature at which the superfluid density of H$_x$TiSe$_2$ vanished in muon-spin rotation measurements\,\cite{Piatti2023CommunPhys}, which suggests that this suppression in $T\ped{c}\apex{A}$ is not a surface effect but rather a bulk property of the material related to vanishing superfluid stiffness.
The absolute values of $\Delta_1$ are rather similar in the two directions, which excludes a large in-plane vs out-of-plane anisotropy of the order parameter.
As for the larger gap $\Delta_2$, it turns out to have a similar amplitude in  $c$-axis and $ab$-plane contacts, always closes at the same $T$ as $\Delta_1$, and its $T$ dependence is fairly similar to a BCS-like one, except for a weak maximum at low $T$ and some deviations at higher $T$ (where, actually, the large gap features are very weak and not well resolved).
Concerning the gap spectral weights, we find $w_1 \approx 0.9$ for $c$ axis contacts and $\approx 0.4$ for $ab$ plane contacts.
Such directional anisotropy is typically expected in superconductors where the Fermi surface cannot be described by a simple sphere\,\cite{Daghero2010SUST, Daghero2011RoPP}.\\

\begin{figure}
	\begin{center}
		\includegraphics[keepaspectratio, width=\columnwidth]{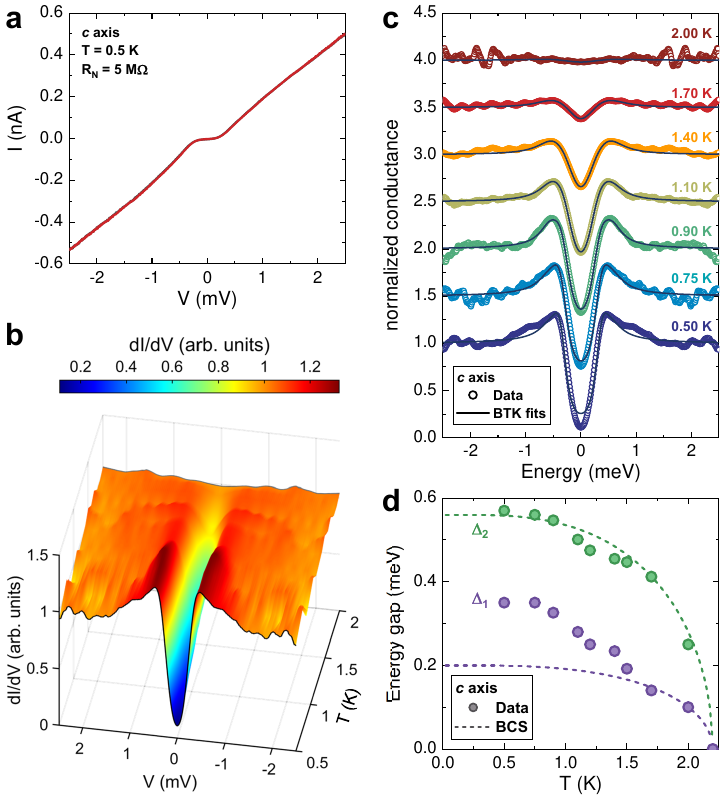}
	\end{center}\vspace{-15pt}
	\caption{
		Tunneling spectroscopy measurements in H$_x$TiSe\ped{2}.
            \textbf{a}, Example of an as-measured current-voltage ($I-V$) curve measured by biasing the sharp Au tip.
            \textbf{b}, Color plot of typical normalized STM tunneling spectra ($dI/dV$ vs. $V$) measured at different temperatures $T$ in zero magnetic field. 
            \textbf{c}, Selected spectra from panel \textbf{b} (symbols) and best-fitting curves (lines) according to the two-band BTK model.
            \textbf{d}, $T$ dependence of the superconducting gaps (symbols) extracted from the fits in panel \textbf{c}. Dashed lines are functions of the form $\Delta_i(T) = \Delta_i(0)\tanh \left(1.74 \sqrt{ T\ped{c}\apex{A}/T-1}\right)$.
	}
	\label{fig:STS}
\end{figure}

\noindent \textbf{Scanning tunneling spectroscopy.}
To further assess the reliability of the PCARS results, we also carried out spectroscopic measurements in the pure tunneling regime, by using a scanning tunnel microscope (STM, see Methods).
Even though the crystals were cleaved just before inserting them in the STM, a residual surface oxidation was detected, preventing good spectroscopic measurements in the scanning mode.
However, the home-built STM we used also allows pushing the sharp Au tip of the microscope against the sample surface, thus piercing the damaged surface layer.
In this configuration, which is actually more similar to a point-contact junction (although with an extremely large $R\ped{N}$, of the order of $5 \times 10^6 \, \Omega$), the $I-V$ curves of the junction display an almost ideal tunnel behavior, as shown in Fig.\ref{fig:STS}a.
Fig.\,\ref{fig:STS}b shows the $T$ dependence of the normalized differential conductance, $d I/dV (V)$, up to 2\,K.
All spectra have been normalized by a linear background fitted to the as-measured data points in a $0.1$\,mV range around $\pm2.5$\,mV.
The spectroscopic signal is very good and the SC gap is clearly visible. Again, in perfect agreement with the results of PCARS, the gap closes around 2\,K.\\
\indent A single-band Dynes fit is actually unable to fit the curves in an accurate way (Supplementary Note\,5). We thus analyzed the same STS spectra by using the identical procedure we used for the PCARS ones, i.e. we fitted the symmetrized normalized conductance curves at various $T$ by using the two-gap, 2D model for Andreev reflection at a S-I-N junction\,\cite{Gonnelli2002PRL, Daghero2010SUST}.
Fig.\,\ref{fig:STS}c shows that the result of the fit is very good at almost all $T$ up to 2\,K. At the lowest $T$, the fitting function is able to reproduce the position and shape of the coherence peaks as well as the shape of the spectra at higher energies, but not the depth of the zero-bias dip.
As a matter of fact, the low-bias shape of the tunnel curves display a change in slope that may suggest either a smaller gap \emph{inside} the coherence peaks, or a broad distribution of gap amplitudes.
The discrepancy is washed out by thermal broadening, so that at 1.10\,K the fit is almost perfectly superimposed to the experimental curve. 
The gap amplitudes extracted from the fit of the $T$-dependent tunnel curves are shown in Fig.\,\ref{fig:STS}d.
The large gap $\Delta_2$ follows rather well a BCS-like $T$ dependence (green dashed line) in the whole $T$ range, whereas the small gap $\Delta_1$ rather displays an almost linear trend that is poorly represented by a standard BCS-like curve.
In particular, if the $\Delta_1(T)$ behavior close to $T\ped{c}$ is thought to agree with a BCS-like curve (purple dashed line), as it happens in the case of PCARS data (see Fig.\,\ref{fig:PCARS}h,l), then the upward linear deviation on cooling starts already at 1.4\,K.
The weight of the small gap is here found to be $w_1\approx0.8$, in reasonable agreement with the PCARS value obtained in $c$ axis contacts -- although it must be noted that the weights do not have equivalent definitions in the Andreev and tunneling regimes\,\cite{Daghero2010SUST, Daghero2011RoPP}.

\section*{Discussion}

\begin{figure*}
	\begin{center}
		\includegraphics[keepaspectratio, width=\textwidth]{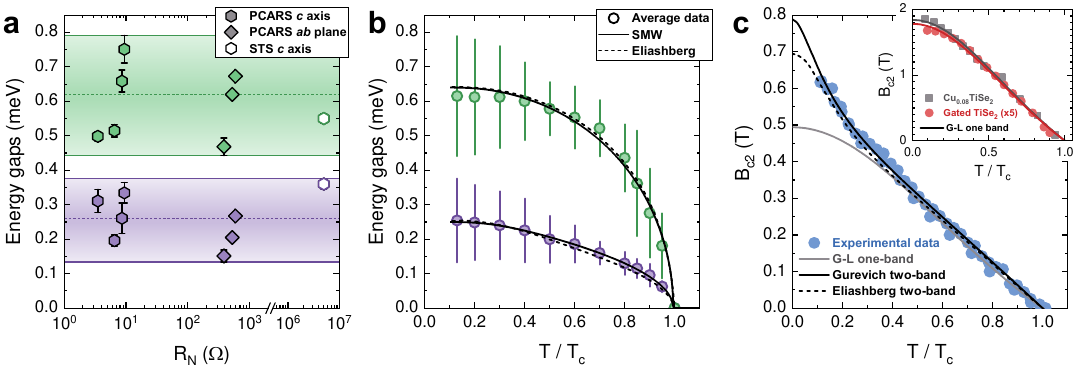}
	\end{center}\vspace{-15pt}
	\caption{
		Two-gap superconductivity in H$_x$TiSe\ped{2}.
            \textbf{a}, Energy gaps measured by PCARS (filled symbols) and STS (hollow symbols) as a function of the contact resistance $R\ped{N}$. Error bars are the spread of gap values obtained by using different normalization choices in the fitting procedure (Methods).
            \textbf{b}, Energy gaps as a function of normalized temperature ($T/T\ped{c}$). Symbols are the mean values obtained by averaging the data measured on all contacts at each $T$. Error bars are the corresponding spread between the different measurements. Solid and dashed lines are the fits with the SMW\,\cite{Suhl1959PRL} and Eliashberg\,\cite{Ummarino2009PRB, Ummarino2004PC, Torsello2019PRB} two-band models.
            \textbf{c}, Upper critical field $B\ped{c2}$ as a function of $T/T\ped{c}$. Symbols are the mean-field data shown in Fig.\,\ref{fig:phase_diagram}. Lines are fits to the one-band Ginzburg-Landau model\,\cite{Tinkham1963PR} (solid gray line), the two-band Gurevich model\,\cite{Gurevich2003PRB, Xing2017SciRep, Ding2022NL} (solid black line) and the two-band Eliashberg model\,\cite{Mansor2005PRB, Ummarino2011JSNM} (dashed black line).
            Inset shows $B\ped{c2}$ vs. $T/T\ped{c}$ for Cu$_x$TiSe\ped{2} (grey squares, Ref.\,\cite{Morosan2006NatPhys}) and ion-gated TiSe\ped{2} (red circles, Ref.\,\cite{Li2016Nature}) and the corresponding fits (solid lines) to the one-band Ginzburg-Landau model\,\cite{Tinkham1963PR}.
	}
	\label{fig:discussion}
\end{figure*}

\noindent Both PCARS and tunneling spectroscopy measurements evidence the two-gap nature of the SC state of H$_x$TiSe\ped{2}.
A summary of the low-$T$ gap amplitudes obtained by PCARS and tunnel spectroscopy is shown in Fig.\,\ref{fig:discussion}a.
It is clearly seen that the gap amplitudes are spread within two well-separated energy ranges, highlighted by shaded bands, and are mostly independent of $R\ped{N}$, supporting the fact that the contacts are in the spectroscopic regime (at least at low $T$).
Using the midpoint and half-width of these ranges as the experimental mean value and uncertainty of the gap amplitudes, one can extract $\langle \Delta_1 \rangle = 0.26 \pm 0.12$ meV and $\langle \Delta_2\rangle = 0.62 \pm 0.18$ meV.
The large spread may indicate a spatial modulation of the gap amplitudes; however, one can exclude that the two gaps pertain to spatially separated regions since:
i) the gap ratio $2\Delta_i/k_BT_c$ is larger than the BCS value for $i=2$ and smaller than the BCS value for $i=1$ (see Fig.\,\ref{fig:PCARS}h and \ref{fig:PCARS}l) and this is what happens in multiple-gap superconductors\,\cite{Daghero2010SUST, Daghero2011RoPP};
ii) STS measurements, that are intrinsically very local in space, display features associated to both the gaps.\\
\begin{table}
    \centering
    \begin{tabular}{ccccccccccccccc}
    \toprule
        & & $\lambda_{11}$ & & $\lambda_{22}$ & & $\lambda_{21}$ & & $\nu_{21}$ & & $\lambda\ped{eff}$ & & $\eta_{21}$ & & $B\ped{c2}(0)$ \\
    \midrule
        SMW + Gurevich & & 0.14 & & 0.22 & & 0.06 & & 0.48 & & 0.34 & & 12.4 & & 0.79\,T \\
        Eliashberg & & 0.24 & & 0.41 & & 0.07 & & 0.48 & & 0.34 & & 10.5 & & 0.69\,T \\
    \bottomrule
    \end{tabular}
    \caption{Electron-phonon coupling matrix elements $\lambda\ped{ij}$, density of states ratio $\nu_{12}$, total effective coupling constant $\lambda\ped{eff}$, diffusivity ratio $\eta_{21}$, and zero-temperature upper critical field $B\ped{c2}(0)$ calculated either with the weak-coupling SMW-Gurevich model or the strong-coupling Eliashberg model.}
    \label{tab:lambdas}
\end{table}
In Fig.\,\ref{fig:discussion}b we plot the $T$ dependence of the mean gap amplitudes (symbols), which we obtained by averaging the values determined from the BTK fits to the $dI/dV$ spectra of all measured PCARS and STS contacts at each normalized temperature $T/T\ped{c}$.
The average amplitudes of the large and small gap respectively display BCS-like and quasi-linear $T$ dependencies, already observed in the spectra of individual contacts, which could explain the lack of saturation at low $T$ detected in the superfluid density by muon spin rotation measurements\,\cite{Piatti2023CommunPhys}.
Despite the spread of the experimental values (again, likely pointing to a spatial modulation of the amplitudes of the two gaps), all spectra measured in both PCARS and STS show that both gaps close at the same $T\ped{c}$ -- whereas separate $T\ped{c}$ for each gap would be expected if the two gaps were to reside on spatially-separated regions.
All of these features can be fully captured by the theories for two-gap superconductivity, as confirmed by the excellent agreement between our data and the fits (solid and dashed lines, respectively) to both the Suhl-Matthias-Walker (SMW) model based on two-band BCS theory\,\cite{Suhl1959PRL} (Supplementary Note\,6) and a more general model based on two-band Eliashberg theory\,\cite{Ummarino2009PRB, Ummarino2004PC, Torsello2019PRB} (Supplementary Note\,7).
The models allow determining the electron-phonon coupling matrix
\begin{center}$\begin{pmatrix}
\lambda_{11} & \lambda_{21}\nu_{21} \\
\lambda_{21} & \lambda_{22}
\end{pmatrix}$
\end{center}
where $\nu_{21}=N_2(0)/N_1(0)$ is the ratio between the density of states at the Fermi level in the two bands.
As summarized in Table\,\ref{tab:lambdas}, both models best fit the experimental data when: i) the maximum coupling is the intraband term in the band associated to the large gap; ii) a finite and non-negligible interband coupling is present; iii) the density of states in the band associated to the large gap is nearly half that of the small gap.
The individual values of the coupling constants cannot be directly compared between the two models since their definitions are not entirely equivalent; however, both models concur in finding a total effective coupling constant $\lambda\ped{eff} = 0.34$, indicating a marginal weak-coupling character for this compound.
\\
\indent The two-band models are able not only to account for the non-BCS, quasi-linear $T$ dependence of the small gap, but also to reproduce the anomalous positive curvature of the $T$ dependence of the upper critical field.
As displayed in Fig.\,\ref{fig:discussion}c, both the Gurevich model based on two-band BCS theory\,\cite{Gurevich2003PRB, Xing2017SciRep, Ding2022NL} (solid black line; see Supplementary Note\,8) and the model based on two-band Eliashberg theory\,\cite{Mansor2005PRB, Ummarino2011JSNM} (black dashed line; see Supplementary Note\,9) are well matched with the $B\ped{c2}\apex{mf}$ data obtained from the magnetotransport measurements in the entire $T$ range.
Notably, this excellent agreement is obtained by using the same coupling parameters that allowed reproducing the $T$ dependence of the gaps in the weak- and strong-coupling models respectively, and both models find comparable values for $B\ped{c2}(0)$ and for $\eta_{21}=D_2/D_1$, the ratio between the diffusivities in the two bands.
These findings set the hydrogen-induced SC state firmly apart from those induced in TiSe\ped{2} by either electric field-driven charge doping\,\cite{Li2019NanoLett} and Cu intercalation\,\cite{Morosan2006NatPhys}, where the $T$ dependences of $B\ped{c2}$ (red and black symbols in the inset to Fig.\,\ref{fig:discussion}c respectively) display the conventional negative curvature commonly observed in superconductors with a single gap, and are well reproduced by the one-band Ginzburg-Landau model (solid lines).\\
\indent Finally, we discuss possible origins for the two-gap SC state observed in TiSe\ped{2} upon H intercalation but not upon other kinds of electron doping.
At low $T$ (and ambient pressure), the Fermi surface of undoped TiSe$_2$ is composed of six equivalent electron pockets located at the L points of the first Brillouin Zone and can thus be described by a single band\,\cite{Knowles2020PRL, Watson2019PRL}.
Moderate electron doping levels, such as those corresponding to the SC phase induced by Cu intercalation, shift the Fermi level to higher energies but do not appreciably alter the band structure\,\cite{Jeong2010PRB, Piatti2023CommunPhys}.
Conversely, the application of external pressure to TiSe$_2$ induces a SC phase in correspondence with an abrupt change in the Fermi surface topology (known as a Lifshitz transition) with the appearance of large additional hole pockets around the $\Gamma$ point\,\cite{Hinlopen2024SciAdv}.
This makes the Fermi surface of pressurized TiSe$_2$ multi-band, with a comparable topology to that of Nb-based SC TMDs (NbS$_2$, NbSe$_2$)\,\cite{Guillamon2008PRL, Kacmarcik2010PC, Yokoya2001Science, Boaknin2003PRL, Guillamon2008PRB, Rahn2012PRB, Noat2015PRB}.
Since no sign of charge compensation due to hole pockets was observed in the Hall effect of H$_x$TiSe$_2$\,\cite{Piatti2023CommunPhys}, the most straightforward interpretation for our results would therefore be that H doping induces a Lifshitz transition in TiSe$_2$ to a Fermi surface composed of multiple inequivalent electron pockets -- similar to the behavior displayed by electron-doped Mo- and W-based semiconducting TMDs (MoS$_2$, WS$_2$, MoSe$_2$, WSe$_2$)\,\cite{Piatti2018NL, Zhang2019NL, Piatti2019JPCM, Ding2022NL}.
This simple interpretation is however at odds with the electronic band structure determined for H$_x$TiSe\ped{2} via ab initio calculations, which predict that a single band crosses the Fermi level in the compound\,\cite{Piatti2023CommunPhys}.\\
\indent An alternative scenario is that the two-gap behavior of H$_x$TiSe$_2$ emerges from an extremely anisotropic distribution of the SC gap function in reciprocal space that is not correlated to the topology of the Fermi surface.
Indeed, such a scenario was recently proposed to be responsible for the multi-gap properties of $2H$-NbSe$_2$\,\cite{Sanna2022npjQM}, distinguishing it from the more conventional two-band behavior of the isostructural compound $2H$-NbS$_2$\,\cite{Heil2017PRL}.
In TiSe$_2$, ab initio calculations showed that H intercalation leads to the filling of highly-coupled states with Ti $d_{z^2}$ orbital character (not accessible with other types of electron doping) and to their hybridization with H-derived orbitals which partially screen phonon perturbations, further enhancing their electron-phonon coupling\,\cite{Piatti2023CommunPhys}.
Since the contribution of the Ti $d_{z^2}$ states to the band structure is highly localized along the $\Gamma$--A line, we expect that this profound alteration of the orbital character of the electronic states at the Fermi level may act as an efficient source of gap anisotropy.
In this picture, the effective two-gap behavior of H$_x$TiSe$_2$ would not be supported by a multi-\textit{band} electronic structure, but rather a multi-\textit{orbital} one enabled by the peculiar effects of H intercalation.
Distinguishing between the different scenarios will require a direct probe of the Fermi surface, making measurements such as quantum oscillations or angle-resolved photoemission spectroscopy highly desirable in this compound.

\section*{Conclusions}

In summary, we have investigated the structure of the superconducting order parameter in H-intercalated TiSe$_2$ by combining three techniques that independently probe the gap structure -- magneto-transport, point-contact spectroscopy, and scanning tunneling spectroscopy measurements -- yielding concording results and revealing the existence of two nodeless gaps.
The spread in the gap values determined in different contacts suggests that their amplitude may be spatially modulated, but the local nature of the tunneling probe excludes that the two gaps might simply originate from phase separation.
Furthermore, both the temperature dependence of gaps and of the upper critical magnetic field can be fitted extremely well by using BCS and Eliashberg models for two-band superconductivity, yielding coupling constants, density of states and diffusivity ratios in very good agreement.
A comparison with the Fermiology of the system on the basis of ab initio calculations suggests that the two-gap structure may emerge from a multi-orbital single-band electronic structure rather than a more conventional multi-band Fermi surface.
These facts further establish the effectiveness of gate-driven H intercalation to synthesize novel superconducting compounds with properties unattainable with conventional doping techniques.
	
\section*{Methods}

\begin{footnotesize}

\noindent \textbf{Ionic liquid gating-induced H intercalation.}
Flake-like 1$T$-TiSe\ped{2} single crystals were purchased from HQ Graphene. Hydrogen intercalation was carried out by immersing the freshly-cleaved crystals (typical size $0.5\times0.5\times0.05$\,mm\apex{3}) in 1-ethyl-3-methylimidazolium tetrafluoroborate (Sigma Aldrich) and applying a gate voltage $V\ped{G}=+3.0$\,V between them and a Pt counter-electrode using an Agilent B2961 power source, for $\approx2$\,h and in ambient conditions. Higher voltages and temperatures were avoided to minimize the risk of undesired electrochemical interactions, such as intercalation of organic ions\,\cite{Piatti2022Nanomat}.
Intercalated crystals were thoroughly rinsed with acetone and ethanol and stored in a vacuum desiccator. Further details can be found in Ref.\,\citenum{Piatti2023CommunPhys}.\\

\noindent \textbf{Electric transport measurements.}
The temperature-dependent resistivity was measured with the standard Van der Pauw method\,\cite{vanderpauw} in a Cryogenic Ltd \apex{3}He insert loaded in a \apex{4}He Oxford Instruments cryostat.
The four-probe voltages were measured via Agilent 34420 nanovoltmeters upon excitation with a constant probe current $\approx 100\,\upmu$A sourced via an Agilent B2912 source-measure unit, which was reversed during each measurement to remove common-mode offsets including thermoelectric voltages.
The magnetic field was applied using a 9\,T SC magnet and always kept orthogonal to the samples' \textit{ab} plane.\\

\noindent \textbf{Point-contact Andreev-reflection spectroscopy.}
PCARS measurements were carried out by means of the soft point-contact method\,\cite{Daghero2010SUST} using the same experimental setup employed in the transport measurements.
Small point-like contacts were realized on the intercalated TiSe\ped{2} crystals by stretching thin Au wires on them so that they touched the freshly-cleaved surface only in a single point, and stabilizing the contact with Ag paste (RS Components) so that the measured conductance results from the parallel of several nanoscopic contacts.
The differential conductance ($dI{/}dV$) spectra of the contacts were acquired by numerical differentiation of the $I{-}V$ (current-voltage) characteristic measured in the pseudo-four-probe configuration\,\cite{Daghero2020PRB, Torsello2022npjQM}.\\

\noindent \textbf{Scanning tunneling spectroscopy.}
The STS measurements were performed in a home-made low temperature STM setup, which was installed on the cold plate of a $^3$He Cryomagnetics Inc system enabling measurements at temperatures down to 0.3\,K and magnetic fields up to 8\,T.
The STM head allows all types of STM measurements and on multiple samples during one cooling. For the measurements we used an Au needle, which was formed by controlled collision on a massive piece of Au placed on the sample holder.
The tunneling conductance spectra $dI{/}dV$ were calculated by numerical differentiation from the $I{-}V$ curves.
Prior to the experiments, the sample surfaces were cleaved by exfoliation and subsequently mounted into the STM head at ambient conditions over a period of about 20\,min.\\

\section*{Data availability}
The data that support the findings of this study are available 
from the authors upon reasonable request.

\end{footnotesize}

\begin{footnotesize}

\section*{Acknowledgments}
E.P., D.D. and R.S.G. acknowledge support from the MIUR PRIN-2017 program (Grant No.2017Z8TS5B -- “Tuning and understanding Quantum phases in 2D materials -- Quantum2D”).
E.P. and R.S.G. also acknowledge funding by the European Union - Next Generation EU as part of the PRIN 2022 PNRR project “Continuous THERmal monitoring with wearable mid-InfraRed sensors” (P2022AHXE5).
This publication is also part of the project PNRR-NGEU which has received funding from the MUR–DM 351/2022.
The work at Slovak Academy of Sciences was supported by Projects APVV-23–0624, VEGA 2/0073/24, COST Action No. CA21144 (SUPERQUMAP) and Slovak Academy of Sciences Project IMPULZ IM-2021–42.
We are thankful to G. Profeta, C. Tresca and G. Lamura for fruitful scientific discussions.

\section*{Author contributions}
E.P., R.S.G. and D.D. conceived the idea.
E.P and D.D. performed the transport measurements.
E.P., G.G., R.S.G. and D.D. performed the point-contact measurements.
F.K., P.Sz. and P.Sa. performed the scanning tunneling spectroscopy measurements.
E.P., G.G., R.S.G. and D.D. analyzed the data.
G.A.U. performed the Eliashberg calculations.
E.P., G.G., R.S.G. and D.D. wrote the manuscript with input from all authors.
R.S.G. directed the project.\\
		
\section*{Competing interests}
The authors declare no competing interests.

\section*{Additional information}
The Supplementary Information 
contains further details on the vortex phase diagram, on the PCAR and STS spectra and their fits, and on the two-band models for the temperature dependence of the gaps and the upper critical field.
	
\end{footnotesize}


\begin{thebibliography}{10}
\expandafter\ifx\csname url\endcsname\relax
  \def\url#1{\texttt{#1}}\fi
\expandafter\ifx\csname urlprefix\endcsname\relax\def\urlprefix{URL }\fi
\providecommand{\bibinfo}[2]{#2}
\providecommand{\eprint}[2][]{\url{#2}}

\bibitem{Wilson1969AP}
\bibinfo{author}{Wilson, J. l.~A.} \& \bibinfo{author}{Yoffe, A.~D.}
\newblock \bibinfo{title}{The transition metal dichalcogenides discussion and
  interpretation of the observed optical, electrical and structural
  properties}.
\newblock \emph{\bibinfo{journal}{Adv. Phys.}} \textbf{\bibinfo{volume}{18}},
  \bibinfo{pages}{193--335} (\bibinfo{year}{1969}).

\bibitem{Shen2008MT}
\bibinfo{author}{Shen, K.~M.} \& \bibinfo{author}{Davis, J.~S.}
\newblock \bibinfo{title}{Cuprate high-t\ped{c} superconductors}.
\newblock \emph{\bibinfo{journal}{Mater. Today}} \textbf{\bibinfo{volume}{11}},
  \bibinfo{pages}{14--21} (\bibinfo{year}{2008}).

\bibitem{Stewart1984RMP}
\bibinfo{author}{Stewart, G.~R.}
\newblock \bibinfo{title}{Heavy-fermion systems}.
\newblock \emph{\bibinfo{journal}{Rev. Mod. Phys.}}
  \textbf{\bibinfo{volume}{56}}, \bibinfo{pages}{755} (\bibinfo{year}{1984}).

\bibitem{Stewart2011RMP}
\bibinfo{author}{Stewart, G.~R.}
\newblock \bibinfo{title}{Superconductivity in iron compounds}.
\newblock \emph{\bibinfo{journal}{Rev. Mod. Phys.}}
  \textbf{\bibinfo{volume}{83}}, \bibinfo{pages}{1589} (\bibinfo{year}{2011}).

\bibitem{Fernandes2022Nature}
\bibinfo{author}{Fernandes, R.~M.} \emph{et~al.}
\newblock \bibinfo{title}{Iron pnictides and chalcogenides: a new paradigm for
  superconductivity}.
\newblock \emph{\bibinfo{journal}{Nature}} \textbf{\bibinfo{volume}{601}},
  \bibinfo{pages}{7891} (\bibinfo{year}{2022}).

\bibitem{Manzeli2017NRM}
\bibinfo{author}{Manzeli, S.}, \bibinfo{author}{Ovchinnikov, D.},
  \bibinfo{author}{Pasquier, D.}, \bibinfo{author}{Yazyev, O.~V.} \&
  \bibinfo{author}{Kis, A.}
\newblock \bibinfo{title}{{2D} transition metal dichalcogenides}.
\newblock \emph{\bibinfo{journal}{Nat. Rev. Mater.}}
  \textbf{\bibinfo{volume}{2}}, \bibinfo{pages}{17033} (\bibinfo{year}{2017}).

\bibitem{Choi2017MT}
\bibinfo{author}{Choi, W.} \emph{et~al.}
\newblock \bibinfo{title}{Recent development of two-dimensional transition
  metal dichalcogenides and their applications}.
\newblock \emph{\bibinfo{journal}{Mater. Today}} \textbf{\bibinfo{volume}{20}},
  \bibinfo{pages}{116--130} (\bibinfo{year}{2017}).

\bibitem{Li2021NatCommun}
\bibinfo{author}{Li, Y.~W.} \emph{et~al.}
\newblock \bibinfo{title}{Observation of topological superconductivity in a
  stoichiometric transition metal dichalcogenide {2$M$-WS\ped{2}}}.
\newblock \emph{\bibinfo{journal}{Nat. Commun.}} \textbf{\bibinfo{volume}{12}},
  \bibinfo{pages}{2874} (\bibinfo{year}{2021}).

\bibitem{Klemm2015PC}
\bibinfo{author}{Klemm, R.~A.}
\newblock \bibinfo{title}{Pristine and intercalated transition metal
  dichalcogenide superconductors}.
\newblock \emph{\bibinfo{journal}{Phys. C: Supercond. Appl.}}
  \textbf{\bibinfo{volume}{514}}, \bibinfo{pages}{86--94}
  (\bibinfo{year}{2015}).

\bibitem{Dresselhaus1987MRSB}
\bibinfo{author}{Dresselhaus, M.~S.}
\newblock \bibinfo{title}{Intercalation in layered materials}.
\newblock \emph{\bibinfo{journal}{MRS Bull.}} \textbf{\bibinfo{volume}{12}},
  \bibinfo{pages}{24--28} (\bibinfo{year}{1987}).

\bibitem{Stark2019AdMa}
\bibinfo{author}{Stark, M.~S.}, \bibinfo{author}{Kuntz, K.~L.},
  \bibinfo{author}{Martens, S.~J.} \& \bibinfo{author}{Warren, S.~C.}
\newblock \bibinfo{title}{Intercalation of layered materials from bulk to
  {2D}}.
\newblock \emph{\bibinfo{journal}{Adv. Mater.}} \textbf{\bibinfo{volume}{31}},
  \bibinfo{pages}{1808213} (\bibinfo{year}{2019}).

\bibitem{Wang2012NatNano}
\bibinfo{author}{Wang, Q.~H.}, \bibinfo{author}{Kalantar-Zadeh, K.},
  \bibinfo{author}{Kis, A.}, \bibinfo{author}{Coleman, J.~N.} \&
  \bibinfo{author}{Strano, M.~S.}
\newblock \bibinfo{title}{Electronics and optoelectronics of two-dimensional
  transition metal dichalcogenides}.
\newblock \emph{\bibinfo{journal}{Nat. Nanotechnol.}}
  \textbf{\bibinfo{volume}{7}}, \bibinfo{pages}{699--712}
  (\bibinfo{year}{2012}).

\bibitem{Geim2013Nature}
\bibinfo{author}{Geim, A.~K.} \& \bibinfo{author}{Grigorieva, I.~V.}
\newblock \bibinfo{title}{Van der {Waals} heterostructures}.
\newblock \emph{\bibinfo{journal}{Nature}} \textbf{\bibinfo{volume}{499}},
  \bibinfo{pages}{419--425} (\bibinfo{year}{2013}).

\bibitem{Fiori2014NatNano}
\bibinfo{author}{Fiori, G.} \emph{et~al.}
\newblock \bibinfo{title}{Electronics based on two-dimensional materials}.
\newblock \emph{\bibinfo{journal}{Nat. Nanotechnol.}}
  \textbf{\bibinfo{volume}{9}}, \bibinfo{pages}{768--779}
  (\bibinfo{year}{2014}).

\bibitem{Voiry2015CSR}
\bibinfo{author}{Voiry, D.}, \bibinfo{author}{Mohite, A.} \&
  \bibinfo{author}{Chhowalla, M.}
\newblock \bibinfo{title}{Phase engineering of transition metal
  dichalcogenides}.
\newblock \emph{\bibinfo{journal}{Chem. Soc. Rev.}}
  \textbf{\bibinfo{volume}{44}}, \bibinfo{pages}{2702--2712}
  (\bibinfo{year}{2015}).

\bibitem{Piatti2021NatElectron}
\bibinfo{author}{Piatti, E.} \emph{et~al.}
\newblock \bibinfo{title}{Charge transport mechanisms in inkjet-printed
  thin-film transistors based on two-dimensional materials}.
\newblock \emph{\bibinfo{journal}{Nat. Electron.}}
  \textbf{\bibinfo{volume}{4}}, \bibinfo{pages}{893--905}
  (\bibinfo{year}{2021}).

\bibitem{DiSalvo1976PRB}
\bibinfo{author}{Di~Salvo, F.~J.}, \bibinfo{author}{Moncton, D.~E.} \&
  \bibinfo{author}{Waszczak, J.~V.}
\newblock \bibinfo{title}{Electronic properties and superlattice formation in
  the semimetal {TiSe\ped{2}}}.
\newblock \emph{\bibinfo{journal}{Phys. Rev. B}} \textbf{\bibinfo{volume}{14}},
  \bibinfo{pages}{4321} (\bibinfo{year}{1976}).

\bibitem{Rasch2008PRL}
\bibinfo{author}{Rasch, J. C.~E.}, \bibinfo{author}{Stemmler, T.},
  \bibinfo{author}{M{\"u}ller, B.}, \bibinfo{author}{Dudy, L.} \&
  \bibinfo{author}{Manzke, R.}
\newblock \bibinfo{title}{{$1T$-TiSe$_2$}: Semimetal or semiconductor?}
\newblock \emph{\bibinfo{journal}{Phys. Rev. Lett.}}
  \textbf{\bibinfo{volume}{101}}, \bibinfo{pages}{237602}
  (\bibinfo{year}{2008}).

\bibitem{Rossnagel2002PRB}
\bibinfo{author}{Rossnagel, K.}, \bibinfo{author}{Kipp, L.} \&
  \bibinfo{author}{Skibowski, M.}
\newblock \bibinfo{title}{Charge-density-wave phase transition in
  {1$T$-TiSe$_2$: Excitonic} insulator versus band-type {Jahn-Teller}
  mechanism}.
\newblock \emph{\bibinfo{journal}{Phys. Rev. B}} \textbf{\bibinfo{volume}{65}},
  \bibinfo{pages}{235101} (\bibinfo{year}{2002}).

\bibitem{Kogar2017Science}
\bibinfo{author}{Kogar, A.} \emph{et~al.}
\newblock \bibinfo{title}{Signatures of exciton condensation in a transition
  metal dichalcogenide}.
\newblock \emph{\bibinfo{journal}{Science}} \textbf{\bibinfo{volume}{358}},
  \bibinfo{pages}{1314--1317} (\bibinfo{year}{2017}).

\bibitem{Knowles2020PRL}
\bibinfo{author}{Knowles, P.} \emph{et~al.}
\newblock \bibinfo{title}{Fermi surface reconstruction and electron dynamics at
  the charge-density-wave transition in {TiSe$_2$}}.
\newblock \emph{\bibinfo{journal}{Phys. Rev. Lett.}}
  \textbf{\bibinfo{volume}{124}}, \bibinfo{pages}{167602}
  (\bibinfo{year}{2020}).

\bibitem{Otto2021SciAdv}
\bibinfo{author}{Otto, M.~R.} \emph{et~al.}
\newblock \bibinfo{title}{Mechanisms of electron-phonon coupling unraveled in
  momentum and time: The case of soft phonons in {TiSe$_2$}}.
\newblock \emph{\bibinfo{journal}{Sci. Adv.}} \textbf{\bibinfo{volume}{7}},
  \bibinfo{pages}{eabf2810} (\bibinfo{year}{2021}).

\bibitem{Lin2022PRL}
\bibinfo{author}{Lin, Z.} \emph{et~al.}
\newblock \bibinfo{title}{Dramatic plasmon response to the charge-density-wave
  gap development in {1$T$-TiSe$_2$}}.
\newblock \emph{\bibinfo{journal}{Phys. Rev. Lett.}}
  \textbf{\bibinfo{volume}{129}}, \bibinfo{pages}{187601}
  (\bibinfo{year}{2022}).

\bibitem{Novko2022PRB}
\bibinfo{author}{Novko, D.}, \bibinfo{author}{Torbatian, Z.} \&
  \bibinfo{author}{Lon{\v{c}}ari{\'c}, I.}
\newblock \bibinfo{title}{Electron correlations rule the phonon-driven
  instability in single-layer {TiSe$_2$}}.
\newblock \emph{\bibinfo{journal}{Phys. Rev. B}}
  \textbf{\bibinfo{volume}{106}}, \bibinfo{pages}{245108}
  (\bibinfo{year}{2022}).

\bibitem{Morosan2006NatPhys}
\bibinfo{author}{Morosan, E.} \emph{et~al.}
\newblock \bibinfo{title}{Superconductivity in {Cu\ped{x}TiSe\ped{2}}}.
\newblock \emph{\bibinfo{journal}{Nat. Phys.}} \textbf{\bibinfo{volume}{2}},
  \bibinfo{pages}{544--550} (\bibinfo{year}{2006}).

\bibitem{Kusmartseva2009PRL}
\bibinfo{author}{Kusmartseva, A.~F.}, \bibinfo{author}{Sipos, B.},
  \bibinfo{author}{Berger, H.}, \bibinfo{author}{Forro, L.} \&
  \bibinfo{author}{Tuti{\v{s}}, E.}
\newblock \bibinfo{title}{Pressure induced superconductivity in pristine
  {1\textit{T}-TiSe\ped{2}}}.
\newblock \emph{\bibinfo{journal}{Phys. Rev. Lett.}}
  \textbf{\bibinfo{volume}{103}}, \bibinfo{pages}{236401}
  (\bibinfo{year}{2009}).

\bibitem{Li2016Nature}
\bibinfo{author}{Li, L.~J.} \emph{et~al.}
\newblock \bibinfo{title}{Controlling many-body states by the electric-field
  effect in a two-dimensional material}.
\newblock \emph{\bibinfo{journal}{Nature}} \textbf{\bibinfo{volume}{529}},
  \bibinfo{pages}{185--189} (\bibinfo{year}{2016}).

\bibitem{Morosan2010PRB}
\bibinfo{author}{Morosan, E.} \emph{et~al.}
\newblock \bibinfo{title}{Multiple electronic transitions and superconductivity
  in {Pd\ped{x}TiSe\ped{2}}}.
\newblock \emph{\bibinfo{journal}{Phys. Rev. B}} \textbf{\bibinfo{volume}{81}},
  \bibinfo{pages}{094524} (\bibinfo{year}{2010}).

\bibitem{Liao2021NatCommun}
\bibinfo{author}{Liao, M.} \emph{et~al.}
\newblock \bibinfo{title}{Coexistence of resistance oscillations and the
  anomalous metal phase in a lithium intercalated {TiSe\ped{2}}
  superconductor}.
\newblock \emph{\bibinfo{journal}{Nat. Commun.}} \textbf{\bibinfo{volume}{12}},
  \bibinfo{pages}{5342} (\bibinfo{year}{2021}).

\bibitem{Piatti2023CommunPhys}
\bibinfo{author}{Piatti, E.} \emph{et~al.}
\newblock \bibinfo{title}{Superconductivity induced by gate-driven hydrogen
  intercalation in the charge-density-wave compound {1$T$-TiSe\ped{2}}}.
\newblock \emph{\bibinfo{journal}{Commun. Phys.}} \textbf{\bibinfo{volume}{6}},
  \bibinfo{pages}{202} (\bibinfo{year}{2023}).

\bibitem{Sato2017JPSJ}
\bibinfo{author}{Sato, K.} \emph{et~al.}
\newblock \bibinfo{title}{New lithium- and diamines-intercalated
  superconductors {Li\ped{x}(C\ped{2}H\ped{8}N\ped{2})\ped{y}TiSe\ped{2}} and
  {Li\ped{x}(C\ped{6}H\ped{16}N\ped{2})\ped{y}TiSe\ped{2}}}.
\newblock \emph{\bibinfo{journal}{J. Phys. Soc. Jpn.}}
  \textbf{\bibinfo{volume}{86}}, \bibinfo{pages}{104701}
  (\bibinfo{year}{2017}).

\bibitem{Li2007PRL}
\bibinfo{author}{Li, S.~Y.}, \bibinfo{author}{Wu, G.}, \bibinfo{author}{Chen,
  X.~H.} \& \bibinfo{author}{Taillefer, L.}
\newblock \bibinfo{title}{Single-gap $s$-wave superconductivity near the
  charge-density-wave quantum critical point in {Cu\ped{x}TiSe\ped{2}}}.
\newblock \emph{\bibinfo{journal}{Phys. Rev. Lett.}}
  \textbf{\bibinfo{volume}{99}}, \bibinfo{pages}{107001}
  (\bibinfo{year}{2007}).

\bibitem{Hillier2010PRB}
\bibinfo{author}{Hillier, A.~D.} \emph{et~al.}
\newblock \bibinfo{title}{Probing the superconducting ground state near the
  charge density wave phase transition in {Cu\ped{0.06}TiSe\ped{2}}}.
\newblock \emph{\bibinfo{journal}{Phys. Rev. B}} \textbf{\bibinfo{volume}{81}},
  \bibinfo{pages}{092507} (\bibinfo{year}{2010}).

\bibitem{Kacmarcik2013PRB}
\bibinfo{author}{Ka{\v{c}}mar{\v{c}}{\'\i}k, J.} \emph{et~al.}
\newblock \bibinfo{title}{Heat capacity of single-crystal
  {Cu\ped{x}TiSe\ped{2}} superconductors}.
\newblock \emph{\bibinfo{journal}{Phys. Rev. B}} \textbf{\bibinfo{volume}{88}},
  \bibinfo{pages}{020507} (\bibinfo{year}{2013}).

\bibitem{Zaberchik2010PRB}
\bibinfo{author}{Zaberchik, M.} \emph{et~al.}
\newblock \bibinfo{title}{Possible evidence of a two-gap structure for the
  {Cu\ped{x}TiSe\ped{2}} superconductor}.
\newblock \emph{\bibinfo{journal}{Phys. Rev. B}} \textbf{\bibinfo{volume}{81}},
  \bibinfo{pages}{220505} (\bibinfo{year}{2010}).

\bibitem{Hinlopen2024SciAdv}
\bibinfo{author}{Hinlopen, R. D.~H.} \emph{et~al.}
\newblock \bibinfo{title}{Lifshitz transition enabling superconducting dome
  around a charge-order critical point}.
\newblock \emph{\bibinfo{journal}{Sci. Adv.}} \textbf{\bibinfo{volume}{10}},
  \bibinfo{pages}{eadl3921} (\bibinfo{year}{2024}).

\bibitem{Prando2023PRMater}
\bibinfo{author}{Prando, G.}, \bibinfo{author}{Piatti, E.},
  \bibinfo{author}{Daghero, D.}, \bibinfo{author}{Gonnelli, R.~S.} \&
  \bibinfo{author}{Carretta, P.}
\newblock \bibinfo{title}{Cluster charge-density-wave glass in
  hydrogen-intercalated {TiSe$_2$}}.
\newblock \emph{\bibinfo{journal}{Phys. Rev. Materials}}
  \textbf{\bibinfo{volume}{7}}, \bibinfo{pages}{094002} (\bibinfo{year}{2023}).

\bibitem{Heil2017PRL}
\bibinfo{author}{Heil, C.} \emph{et~al.}
\newblock \bibinfo{title}{Origin of superconductivity and latent charge density
  wave in {NbS$_2$}}.
\newblock \emph{\bibinfo{journal}{Phys. Rev. Lett.}}
  \textbf{\bibinfo{volume}{119}}, \bibinfo{pages}{087003}
  (\bibinfo{year}{2017}).

\bibitem{Sanna2022npjQM}
\bibinfo{author}{Sanna, A.} \emph{et~al.}
\newblock \bibinfo{title}{Real-space anisotropy of the superconducting gap in
  the charge-density wave material {2$H$-NbSe$_2$}}.
\newblock \emph{\bibinfo{journal}{npj Quantum Mater.}}
  \textbf{\bibinfo{volume}{7}}, \bibinfo{pages}{6} (\bibinfo{year}{2022}).

\bibitem{Guillamon2008PRL}
\bibinfo{author}{Guillam{\'o}n, I.} \emph{et~al.}
\newblock \bibinfo{title}{Superconducting density of states and vortex cores of
  {$2H$-NbS$_2$}}.
\newblock \emph{\bibinfo{journal}{Phys. Rev. Lett.}}
  \textbf{\bibinfo{volume}{101}}, \bibinfo{pages}{166407}
  (\bibinfo{year}{2008}).

\bibitem{Kacmarcik2010PC}
\bibinfo{author}{Ka{\v{c}}mar{\v{c}}{\'\i}k, J.} \emph{et~al.}
\newblock \bibinfo{title}{Studies on two-gap superconductivity in
  {$2H$-NbS$_2$}}.
\newblock \emph{\bibinfo{journal}{Phys. C: Supercond. Appl.}}
  \textbf{\bibinfo{volume}{470}}, \bibinfo{pages}{S719--S720}
  (\bibinfo{year}{2010}).

\bibitem{Yokoya2001Science}
\bibinfo{author}{Yokoya, T.} \emph{et~al.}
\newblock \bibinfo{title}{Fermi surface sheet-dependent superconductivity in
  {$2H$-NbSe$_2$}}.
\newblock \emph{\bibinfo{journal}{Science}} \textbf{\bibinfo{volume}{294}},
  \bibinfo{pages}{2518--2520} (\bibinfo{year}{2001}).

\bibitem{Boaknin2003PRL}
\bibinfo{author}{Boaknin, E.} \emph{et~al.}
\newblock \bibinfo{title}{Heat conduction in the vortex state of {NbSe$_2$}:
  {Evidence} for multiband superconductivity}.
\newblock \emph{\bibinfo{journal}{Phys. Rev. Lett.}}
  \textbf{\bibinfo{volume}{90}}, \bibinfo{pages}{117003}
  (\bibinfo{year}{2003}).

\bibitem{Guillamon2008PRB}
\bibinfo{author}{Guillamon, I.}, \bibinfo{author}{Suderow, H.},
  \bibinfo{author}{Guinea, F.} \& \bibinfo{author}{Vieira, S.}
\newblock \bibinfo{title}{Intrinsic atomic-scale modulations of the
  superconducting gap of {$2H$-NbSe$_2$}}.
\newblock \emph{\bibinfo{journal}{Phys. Rev. B}} \textbf{\bibinfo{volume}{77}},
  \bibinfo{pages}{134505} (\bibinfo{year}{2008}).

\bibitem{Rahn2012PRB}
\bibinfo{author}{Rahn, D.~J.} \emph{et~al.}
\newblock \bibinfo{title}{Gaps and kinks in the electronic structure of the
  superconductor {2$H$-NbSe$_2$} from angle-resolved photoemission at {1\,K}}.
\newblock \emph{\bibinfo{journal}{Phys. Rev. B}} \textbf{\bibinfo{volume}{85}},
  \bibinfo{pages}{224532} (\bibinfo{year}{2012}).

\bibitem{Noat2015PRB}
\bibinfo{author}{Noat, Y.} \emph{et~al.}
\newblock \bibinfo{title}{Quasiparticle spectra of {$2H$-NbSe$_2$}: {Two}-band
  superconductivity and the role of tunneling selectivity}.
\newblock \emph{\bibinfo{journal}{Phys. Rev. B}} \textbf{\bibinfo{volume}{92}},
  \bibinfo{pages}{134510} (\bibinfo{year}{2015}).

\bibitem{Boeri2022JPCM}
\bibinfo{author}{Boeri, L.} \emph{et~al.}
\newblock \bibinfo{title}{The 2021 room-temperature superconductivity roadmap}.
\newblock \emph{\bibinfo{journal}{J. Phys. Condens. Matter}}
  \textbf{\bibinfo{volume}{34}}, \bibinfo{pages}{183002}
  (\bibinfo{year}{2022}).

\bibitem{Lu2017Nature}
\bibinfo{author}{Lu, N.} \emph{et~al.}
\newblock \bibinfo{title}{Electric-field control of tri-state phase
  transformation with a selective dual-ion switch}.
\newblock \emph{\bibinfo{journal}{Nature}} \textbf{\bibinfo{volume}{546}},
  \bibinfo{pages}{124--128} (\bibinfo{year}{2017}).

\bibitem{Meng2022PRB}
\bibinfo{author}{Meng, Y.} \emph{et~al.}
\newblock \bibinfo{title}{Protonation-induced discrete superconducting phases
  in bulk {FeSe} single crystals}.
\newblock \emph{\bibinfo{journal}{Phys. Rev. B}}
  \textbf{\bibinfo{volume}{105}}, \bibinfo{pages}{134506}
  (\bibinfo{year}{2022}).

\bibitem{Saito2018NatCommun}
\bibinfo{author}{Saito, Y.}, \bibinfo{author}{Nojima, T.} \&
  \bibinfo{author}{Iwasa, Y.}
\newblock \bibinfo{title}{Quantum phase transitions in highly crystalline
  two-dimensional superconductors}.
\newblock \emph{\bibinfo{journal}{Nat. Commun.}} \textbf{\bibinfo{volume}{9}},
  \bibinfo{pages}{778} (\bibinfo{year}{2018}).

\bibitem{Li2019NanoLett}
\bibinfo{author}{Li, L.} \emph{et~al.}
\newblock \bibinfo{title}{Anomalous quantum metal in a {2D} crystalline
  superconductor with electronic phase nonuniformity}.
\newblock \emph{\bibinfo{journal}{Nano Lett.}} \textbf{\bibinfo{volume}{19}},
  \bibinfo{pages}{4126--4133} (\bibinfo{year}{2019}).

\bibitem{Altshuler1980}
\bibinfo{author}{Altshuler, B.~L.}, \bibinfo{author}{Aronov, A.~G.} \&
  \bibinfo{author}{Lee, P.~A.}
\newblock \bibinfo{title}{Interaction effects in disordered {Fermi} systems in
  two dimensions}.
\newblock \emph{\bibinfo{journal}{Phys. Rev. Lett.}}
  \textbf{\bibinfo{volume}{44}}, \bibinfo{pages}{1288} (\bibinfo{year}{1980}).

\bibitem{FukuyamaBook}
\bibinfo{author}{Fukuyama, H.}
\newblock \emph{\bibinfo{title}{Electron–electron Interactions in Disordered
  Systems}} (\bibinfo{publisher}{North Holland}, \bibinfo{address}{Amsterdam},
  \bibinfo{year}{1985}).

\bibitem{Bianco2015PRB}
\bibinfo{author}{Bianco, R.}, \bibinfo{author}{Calandra, M.} \&
  \bibinfo{author}{Mauri, F.}
\newblock \bibinfo{title}{Electronic and vibrational properties of
  {TiSe\ped{2}} in the charge-density-wave phase from first principles}.
\newblock \emph{\bibinfo{journal}{Phys. Rev. B}} \textbf{\bibinfo{volume}{92}},
  \bibinfo{pages}{094107} (\bibinfo{year}{2015}).

\bibitem{Rohwer2011}
\bibinfo{author}{Rohwer, T.} \emph{et~al.}
\newblock \bibinfo{title}{Collapse of long-range charge order tracked by
  time-resolved photoemission at high momenta}.
\newblock \emph{\bibinfo{journal}{Nature}} \textbf{\bibinfo{volume}{471}},
  \bibinfo{pages}{490--493} (\bibinfo{year}{2011}).

\bibitem{Blatter1994RMP}
\bibinfo{author}{Blatter, G.}, \bibinfo{author}{Feigel'man, M.~V.},
  \bibinfo{author}{Geshkenbein, V.~B.}, \bibinfo{author}{Larkin, A.~I.} \&
  \bibinfo{author}{Vinokur, V.~M.}
\newblock \bibinfo{title}{Vortices in high-temperature superconductors}.
\newblock \emph{\bibinfo{journal}{Rev. Mod. Phys.}}
  \textbf{\bibinfo{volume}{66}}, \bibinfo{pages}{1125} (\bibinfo{year}{1994}).

\bibitem{Lewellyn2019PRB}
\bibinfo{author}{Lewellyn, N.~A.} \emph{et~al.}
\newblock \bibinfo{title}{Infinite-randomness fixed point of the quantum
  superconductor-metal transitions in amorphous thin films}.
\newblock \emph{\bibinfo{journal}{Phys. Rev. B}} \textbf{\bibinfo{volume}{99}},
  \bibinfo{pages}{054515} (\bibinfo{year}{2019}).

\bibitem{Wang2023PRB}
\bibinfo{author}{Wang, X.} \emph{et~al.}
\newblock \bibinfo{title}{Robust quantum griffiths singularity above 1.5 k in
  nitride thin films}.
\newblock \emph{\bibinfo{journal}{Phys. Rev. B}}
  \textbf{\bibinfo{volume}{107}}, \bibinfo{pages}{094509}
  (\bibinfo{year}{2023}).

\bibitem{Theunissen1997PRB}
\bibinfo{author}{Theunissen, M.~H.} \& \bibinfo{author}{Kes, P.~H.}
\newblock \bibinfo{title}{Resistive transitions of thin film superconductors in
  a magnetic field}.
\newblock \emph{\bibinfo{journal}{Phys. Rev. B}} \textbf{\bibinfo{volume}{55}},
  \bibinfo{pages}{15183} (\bibinfo{year}{1997}).

\bibitem{Berghuis1993PRB}
\bibinfo{author}{Berghuis, P.} \& \bibinfo{author}{Kes, P.~H.}
\newblock \bibinfo{title}{Two-dimensional collective pinning and vortex-lattice
  melting in {$a$-Nb\ped{1-x}Ge\ped{x} films}}.
\newblock \emph{\bibinfo{journal}{Phys. Rev. B}} \textbf{\bibinfo{volume}{47}},
  \bibinfo{pages}{262} (\bibinfo{year}{1993}).

\bibitem{Tinkham1963PR}
\bibinfo{author}{Tinkham, M.}
\newblock \bibinfo{title}{Effect of fluxoid quantization on transitions of
  superconducting films}.
\newblock \emph{\bibinfo{journal}{Phys. Rev.}} \textbf{\bibinfo{volume}{129}},
  \bibinfo{pages}{2413} (\bibinfo{year}{1963}).

\bibitem{Ullah1990PRL}
\bibinfo{author}{Ullah, S.} \& \bibinfo{author}{Dorsey, A.~T.}
\newblock \bibinfo{title}{Critical fluctuations in high-temperature
  superconductors and the {Ettingshausen} effect}.
\newblock \emph{\bibinfo{journal}{Phys. Rev. Lett.}}
  \textbf{\bibinfo{volume}{65}}, \bibinfo{pages}{2066} (\bibinfo{year}{1990}).

\bibitem{Dezi2018PRB}
\bibinfo{author}{Dezi, G.}, \bibinfo{author}{Scopigno, N.},
  \bibinfo{author}{Caprara, S.} \& \bibinfo{author}{Grilli, M.}
\newblock \bibinfo{title}{Negative electronic compressibility and nanoscale
  inhomogeneity in ionic-liquid gated two-dimensional superconductors}.
\newblock \emph{\bibinfo{journal}{Phys. Rev. B}} \textbf{\bibinfo{volume}{98}},
  \bibinfo{pages}{214507} (\bibinfo{year}{2018}).

\bibitem{Venditti2019PRB}
\bibinfo{author}{Venditti, G.} \emph{et~al.}
\newblock \bibinfo{title}{Nonlinear {I-V} characteristics of two-dimensional
  superconductors: {Berezinskii-Kosterlitz-Thouless} physics versus
  inhomogeneity}.
\newblock \emph{\bibinfo{journal}{Phys. Rev. B}}
  \textbf{\bibinfo{volume}{100}}, \bibinfo{pages}{064506}
  (\bibinfo{year}{2019}).

\bibitem{Xing2021NanoLett}
\bibinfo{author}{Xing, Y.} \emph{et~al.}
\newblock \bibinfo{title}{Extrinsic and intrinsic anomalous metallic states in
  transition metal dichalcogenide {Ising} superconductors}.
\newblock \emph{\bibinfo{journal}{Nano Lett.}} \textbf{\bibinfo{volume}{21}},
  \bibinfo{pages}{7486--7494} (\bibinfo{year}{2021}).

\bibitem{Gurevich2003PRB}
\bibinfo{author}{Gurevich, A.}
\newblock \bibinfo{title}{Enhancement of the upper critical field by
  nonmagnetic impurities in dirty two-gap superconductors}.
\newblock \emph{\bibinfo{journal}{Phys. Rev. B}} \textbf{\bibinfo{volume}{67}},
  \bibinfo{pages}{184515} (\bibinfo{year}{2003}).

\bibitem{Xing2017SciRep}
\bibinfo{author}{Xing, X.} \emph{et~al.}
\newblock \bibinfo{title}{Two-band and {Pauli}-limiting effects on the upper
  critical field of 112-type iron pnictide superconductors}.
\newblock \emph{\bibinfo{journal}{Sci. Rep.}} \textbf{\bibinfo{volume}{7}},
  \bibinfo{pages}{45943} (\bibinfo{year}{2017}).

\bibitem{Ding2022NL}
\bibinfo{author}{Ding, D.} \emph{et~al.}
\newblock \bibinfo{title}{Multivalley superconductivity in monolayer transition
  metal dichalcogenides}.
\newblock \emph{\bibinfo{journal}{Nano Lett.}} \textbf{\bibinfo{volume}{22}},
  \bibinfo{pages}{7919--7926} (\bibinfo{year}{2022}).

\bibitem{Naidyuk2005Book}
\bibinfo{author}{Naidyuk, Y.~G.} \& \bibinfo{author}{Yanson, I.~K.}
\newblock \emph{\bibinfo{title}{Point-contact spectroscopy}}
  (\bibinfo{publisher}{Springer New York, NY}, \bibinfo{year}{2005}).

\bibitem{Daghero2010SUST}
\bibinfo{author}{Daghero, D.} \& \bibinfo{author}{Gonnelli, R.~S.}
\newblock \bibinfo{title}{Probing multiband superconductivity by point-contact
  spectroscopy}.
\newblock \emph{\bibinfo{journal}{Supercond. Sci. Technol.}}
  \textbf{\bibinfo{volume}{23}}, \bibinfo{pages}{043001}
  (\bibinfo{year}{2010}).

\bibitem{Torsello2022npjQM}
\bibinfo{author}{Torsello, D.} \emph{et~al.}
\newblock \bibinfo{title}{Nodal multigap superconductivity in the anisotropic
  iron-based compound {RbCa$_2$Fe$_4$As$_4$F$_2$}}.
\newblock \emph{\bibinfo{journal}{npj Quantum Materials}}
  \textbf{\bibinfo{volume}{7}}, \bibinfo{pages}{1--7} (\bibinfo{year}{2022}).

\bibitem{Gonnelli2002PRL}
\bibinfo{author}{Gonnelli, R.~S.} \emph{et~al.}
\newblock \bibinfo{title}{Direct evidence for two-band superconductivity in
  {MgB$_2$} single crystals from directional point-contact spectroscopy in
  magnetic fields}.
\newblock \emph{\bibinfo{journal}{Phys. Rev. Lett.}}
  \textbf{\bibinfo{volume}{89}}, \bibinfo{pages}{247004}
  (\bibinfo{year}{2002}).

\bibitem{Gonnelli2004PRB}
\bibinfo{author}{Gonnelli, R.~S.} \emph{et~al.}
\newblock \bibinfo{title}{Magnetic-field dependence of the gaps in a two-band
  superconductor: A point-contact study of {MgB$_2$} single crystals}.
\newblock \emph{\bibinfo{journal}{Phys. Rev. B}} \textbf{\bibinfo{volume}{69}},
  \bibinfo{pages}{100504(R)} (\bibinfo{year}{2004}).

\bibitem{Daghero2011RoPP}
\bibinfo{author}{Daghero, D.}, \bibinfo{author}{Tortello, M.},
  \bibinfo{author}{Ummarino, G.~A.} \& \bibinfo{author}{Gonnelli, R.~S.}
\newblock \bibinfo{title}{Directional point-contact {Andreev}-reflection
  spectroscopy of fe-based superconductors: {Fermi} surface topology, gap
  symmetry, and electron--boson interaction}.
\newblock \emph{\bibinfo{journal}{Rep. Prog. Phys.}}
  \textbf{\bibinfo{volume}{74}}, \bibinfo{pages}{124509}
  (\bibinfo{year}{2011}).

\bibitem{Daghero2020PRB}
\bibinfo{author}{Daghero, D.} \emph{et~al.}
\newblock \bibinfo{title}{Superconductivity of underdoped {PrFeAs(O,F)}
  investigated via point-contact spectroscopy and nuclear magnetic resonance}.
\newblock \emph{\bibinfo{journal}{Phys. Rev. B}}
  \textbf{\bibinfo{volume}{102}}, \bibinfo{pages}{104513}
  (\bibinfo{year}{2020}).

\bibitem{Daghero2023LTP}
\bibinfo{author}{Daghero, D.}, \bibinfo{author}{Piatti, E.},
  \bibinfo{author}{Zhigadlo, N.~D.} \& \bibinfo{author}{Gonnelli, R.~S.}
\newblock \bibinfo{title}{A model for critical current effects in point-contact
  {Andreev}-reflection spectroscopy}.
\newblock \emph{\bibinfo{journal}{Low Temp. Phys.}}
  \textbf{\bibinfo{volume}{49}}, \bibinfo{pages}{886--892}
  (\bibinfo{year}{2023}).

\bibitem{Sheet2004}
\bibinfo{author}{Sheet, G.}, \bibinfo{author}{Mukhopadhyay, S.} \&
  \bibinfo{author}{Raychaudhuri, P.}
\newblock \bibinfo{title}{Role of critical current on the point-contact
  {Andreev} reflection spectra between a normal metal and a superconductor}.
\newblock \emph{\bibinfo{journal}{Phys. Rev. B}} \textbf{\bibinfo{volume}{69}},
  \bibinfo{pages}{134507} (\bibinfo{year}{2004}).

\bibitem{Chen2010PRB}
\bibinfo{author}{Chen, T.~Y.}, \bibinfo{author}{Huang, S.~X.} \&
  \bibinfo{author}{Chien, C.~L.}
\newblock \bibinfo{title}{Pronounced effects of additional resistance in
  {Andreev} reflection spectroscopy}.
\newblock \emph{\bibinfo{journal}{Phys. Rev. B}} \textbf{\bibinfo{volume}{81}},
  \bibinfo{pages}{214444} (\bibinfo{year}{2010}).

\bibitem{Doring2014}
\bibinfo{author}{D{\"o}ring, S.} \emph{et~al.}
\newblock \bibinfo{title}{Influence of the spreading resistance on the
  conductance spectrum of planar hybrid thin film {SNS'} junctions based on
  iron pnictides}.
\newblock \emph{\bibinfo{journal}{J. Phys. Conf. Ser.}}
  \textbf{\bibinfo{volume}{507}}, \bibinfo{pages}{012008}
  (\bibinfo{year}{2014}).

\bibitem{Piatti2024Nanomat}
\bibinfo{author}{Piatti, E.} \emph{et~al.}
\newblock \bibinfo{title}{{Superconductivity of Co-Doped CaKFe$_4$As$_4$
  Investigated via Point-Contact Spectroscopy and London Penetration Depth
  Measurements}}.
\newblock \emph{\bibinfo{journal}{Nanomaterials}}
  \textbf{\bibinfo{volume}{14}}, \bibinfo{pages}{1319} (\bibinfo{year}{2024}).

\bibitem{Kashiwaya2000RoPP}
\bibinfo{author}{Kashiwaya, S.} \& \bibinfo{author}{Tanaka, Y.}
\newblock \bibinfo{title}{Tunnelling effects on surface bound states in
  unconventional superconductors}.
\newblock \emph{\bibinfo{journal}{Rep. Prog. Phys.}}
  \textbf{\bibinfo{volume}{63}}, \bibinfo{pages}{1641} (\bibinfo{year}{2000}).
\newblock \urlprefix\url{https://dx.doi.org/10.1088/0034-4885/63/10/202}.

\bibitem{KashiwayaPRB1996}
\bibinfo{author}{Kashiwaya, S.}, \bibinfo{author}{Tanaka, Y.},
  \bibinfo{author}{Koyanagi, M.} \& \bibinfo{author}{Kajimura, K.}
\newblock \bibinfo{title}{Theory for tunneling spectroscopy of anisotropic
  superconductors}.
\newblock \emph{\bibinfo{journal}{Phys. Rev. B}} \textbf{\bibinfo{volume}{53}},
  \bibinfo{pages}{2667--2676} (\bibinfo{year}{1996}).

\bibitem{BlonderPRB1982}
\bibinfo{author}{Blonder, G.~E.}, \bibinfo{author}{Tinkham, M.} \&
  \bibinfo{author}{Klapwijk, T.~M.}
\newblock \bibinfo{title}{Transition from metallic to tunneling regimes in
  superconducting microconstrictions: Excess current, charge imbalance, and
  supercurrent conversion}.
\newblock \emph{\bibinfo{journal}{Phys. Rev. B}} \textbf{\bibinfo{volume}{25}},
  \bibinfo{pages}{4515--4532} (\bibinfo{year}{1982}).

\bibitem{zhao_2019NaturePhys}
\bibinfo{author}{Zhao, K.} \emph{et~al.}
\newblock \bibinfo{title}{Disorder-induced multifractal superconductivity in
  monolayer niobium dichalcogenides}.
\newblock \emph{\bibinfo{journal}{Nat. Phys.}} \textbf{\bibinfo{volume}{15}},
  \bibinfo{pages}{904--910} (\bibinfo{year}{2019}).

\bibitem{Suhl1959PRL}
\bibinfo{author}{Suhl, H.}, \bibinfo{author}{Matthias, B.} \&
  \bibinfo{author}{Walker, L.~R.}
\newblock \bibinfo{title}{{Bardeen-Cooper-Schrieffer theory of
  superconductivity in the case of overlapping bands}}.
\newblock \emph{\bibinfo{journal}{Phys. Rev. Lett.}}
  \textbf{\bibinfo{volume}{3}}, \bibinfo{pages}{552} (\bibinfo{year}{1959}).

\bibitem{Ummarino2009PRB}
\bibinfo{author}{Ummarino, G.~A.}, \bibinfo{author}{Tortello, M.},
  \bibinfo{author}{Daghero, D.} \& \bibinfo{author}{Gonnelli, R.~S.}
\newblock \bibinfo{title}{{Three-band s$\pm$Eliashberg theory and the
  superconducting gaps of iron pnictides}}.
\newblock \emph{\bibinfo{journal}{Phys. Rev. B}} \textbf{\bibinfo{volume}{80}},
  \bibinfo{pages}{172503} (\bibinfo{year}{2009}).

\bibitem{Ummarino2004PC}
\bibinfo{author}{Ummarino, G.~A.}, \bibinfo{author}{Gonnelli, R.~S.},
  \bibinfo{author}{Massidda, S.} \& \bibinfo{author}{Bianconi, A.}
\newblock \bibinfo{title}{{Two-band Eliashberg equations and the experimental
  T\ped{c} of the diboride Mg$_{1-x}$Al$_x$B$_2$}}.
\newblock \emph{\bibinfo{journal}{Phys. C: Supercond. Appl.}}
  \textbf{\bibinfo{volume}{407}}, \bibinfo{pages}{121--127}
  (\bibinfo{year}{2004}).

\bibitem{Torsello2019PRB}
\bibinfo{author}{Torsello, D.} \emph{et~al.}
\newblock \bibinfo{title}{{Analysis of the London penetration depth in Ni-doped
  CaKFe$_4$As$_4$}}.
\newblock \emph{\bibinfo{journal}{Phys. Rev. B}}
  \textbf{\bibinfo{volume}{100}}, \bibinfo{pages}{094513}
  (\bibinfo{year}{2019}).

\bibitem{Mansor2005PRB}
\bibinfo{author}{Mansor, M.} \& \bibinfo{author}{Carbotte, J.~P.}
\newblock \bibinfo{title}{Upper critical field in two-band superconductivity}.
\newblock \emph{\bibinfo{journal}{Phys. Rev. B}} \textbf{\bibinfo{volume}{72}},
  \bibinfo{pages}{024538} (\bibinfo{year}{2005}).

\bibitem{Ummarino2011JSNM}
\bibinfo{author}{Ummarino, G.~A.}, \bibinfo{author}{Daghero, D.},
  \bibinfo{author}{Tortello, M.} \& \bibinfo{author}{Gonnelli, R.~S.}
\newblock \bibinfo{title}{{Predictions of multiband s$\pm$strong-coupling
  Eliashberg theory compared to experimental data in iron pnictides}}.
\newblock \emph{\bibinfo{journal}{J. Supecond. Novel Magn.}}
  \textbf{\bibinfo{volume}{24}}, \bibinfo{pages}{247--253}
  (\bibinfo{year}{2011}).

\bibitem{Watson2019PRL}
\bibinfo{author}{Watson, M.~D.} \emph{et~al.}
\newblock \bibinfo{title}{{Orbital-and $k_z$-selective hybridization of Se $4p$
  and Ti $3d$ states in the charge density wave phase of TiSe$_2$}}.
\newblock \emph{\bibinfo{journal}{Phys. Rev. Lett.}}
  \textbf{\bibinfo{volume}{122}}, \bibinfo{pages}{076404}
  (\bibinfo{year}{2019}).

\bibitem{Jeong2010PRB}
\bibinfo{author}{Jeong, J.}, \bibinfo{author}{Jeong, J.}, \bibinfo{author}{Noh,
  H.-J.}, \bibinfo{author}{Kim, S.~B.} \& \bibinfo{author}{Kim, H.-D.}
\newblock \bibinfo{title}{{Electronic structure study of Cu-doped 1T-TiSe$_2$
  by angle-resolved photoemission spectroscopy}}.
\newblock \emph{\bibinfo{journal}{Phys. C: Supercond. Appl.}}
  \textbf{\bibinfo{volume}{470}}, \bibinfo{pages}{S648--S650}
  (\bibinfo{year}{2010}).

\bibitem{Piatti2018NL}
\bibinfo{author}{Piatti, E.} \emph{et~al.}
\newblock \bibinfo{title}{Multi-valley superconductivity in ion-gated mos$_2$
  layers}.
\newblock \emph{\bibinfo{journal}{Nano Lett.}} \textbf{\bibinfo{volume}{18}},
  \bibinfo{pages}{4821--4830} (\bibinfo{year}{2018}).

\bibitem{Zhang2019NL}
\bibinfo{author}{Zhang, H.}, \bibinfo{author}{Berthod, C.},
  \bibinfo{author}{Berger, H.}, \bibinfo{author}{Giamarchi, T.} \&
  \bibinfo{author}{Morpurgo, A.~F.}
\newblock \bibinfo{title}{Band filling and cross quantum capacitance in
  ion-gated semiconducting transition metal dichalcogenide monolayers}.
\newblock \emph{\bibinfo{journal}{Nano Lett.}} \textbf{\bibinfo{volume}{19}},
  \bibinfo{pages}{8836--8845} (\bibinfo{year}{2019}).

\bibitem{Piatti2019JPCM}
\bibinfo{author}{Piatti, E.}, \bibinfo{author}{Romanin, D.} \&
  \bibinfo{author}{Gonnelli, R.~S.}
\newblock \bibinfo{title}{{Mapping multi-valley Lifshitz transitions induced by
  field-effect doping in strained MoS$_2$ nanolayers}}.
\newblock \emph{\bibinfo{journal}{J. Phys. Condens. Matter}}
  \textbf{\bibinfo{volume}{31}}, \bibinfo{pages}{114002}
  (\bibinfo{year}{2019}).

\bibitem{Piatti2022Nanomat}
\bibinfo{author}{Piatti, E.}, \bibinfo{author}{Montagna~Bozzone, J.} \&
  \bibinfo{author}{Daghero, D.}
\newblock \bibinfo{title}{Anomalous metallic phase in molybdenum disulphide
  induced via gate-driven organic ion intercalation}.
\newblock \emph{\bibinfo{journal}{Nanomaterials}}
  \textbf{\bibinfo{volume}{12}}, \bibinfo{pages}{1842} (\bibinfo{year}{2022}).

\bibitem{vanderpauw}
\bibinfo{author}{Lim, S. H.~N.}, \bibinfo{author}{McKenzie, D.~R.} \&
  \bibinfo{author}{Bilek, M. M.~M.}
\newblock \bibinfo{title}{Van der {Pauw} method for measuring resistivity of a
  plane sample with distant boundaries}.
\newblock \emph{\bibinfo{journal}{Rev. Sci. Instrum.}}
  \textbf{\bibinfo{volume}{80}} (\bibinfo{year}{2009}).

\end{thebibliography}
\end{document}


\title{Supplementary Information for\texorpdfstring{\\}{ } 
Direct evidence for two-gap superconductivity in hydrogen-intercalated titanium diselenide
}
	
\author{Erik Piatti}
\email{erik.piatti@polito.it}
\affiliation{Department of Applied Science and Technology, Politecnico di Torino, I-10129 Torino, Italy}
\author{Gaia Gavello}
\affiliation{Department of Applied Science and Technology, Politecnico di Torino, I-10129 Torino, Italy}
\author{Giovanni A. Ummarino}
\affiliation{Department of Applied Science and Technology, Politecnico di Torino, I-10129 Torino, Italy}
\author{Filip Ko{\v s}uth}
\affiliation{Centre of Low Temperature Physics, Institute of Experimental Physics, Slovak Academy of Sciences, SK-04001 Ko{\v s}ice, Slovakia}
\affiliation{Centre of Low Temperature Physics, Faculty of Science, P. J. {\v S}af{\'a}rik University, SK-04001 Ko{\v s}ice, Slovakia}
\author{Pavol Szab{\'o}}
\author{Peter Samuely}
\affiliation{Centre of Low Temperature Physics, Institute of Experimental Physics, Slovak Academy of Sciences, SK-04001 Ko{\v s}ice, Slovakia}
\author{Renato S. Gonnelli}
\email{renato.gonnelli@polito.it}
\affiliation{Department of Applied Science and Technology, Politecnico di Torino, I-10129 Torino, Italy}
\author{Dario Daghero}
\affiliation{Department of Applied Science and Technology, Politecnico di Torino, I-10129 Torino, Italy}

\maketitle

\renewcommand{\theequation}{S\arabic{equation}}
\renewcommand{\citenumfont}[1]{S#1} 
\renewcommand{\bibnumfmt}[1]{[S#1]} 



\widetext

\section*{Supplementary Note 1: Further details on the $T-B$ phase diagram} 

\begin{figure*}
    \begin{center}
		\includegraphics[keepaspectratio, width=\textwidth]{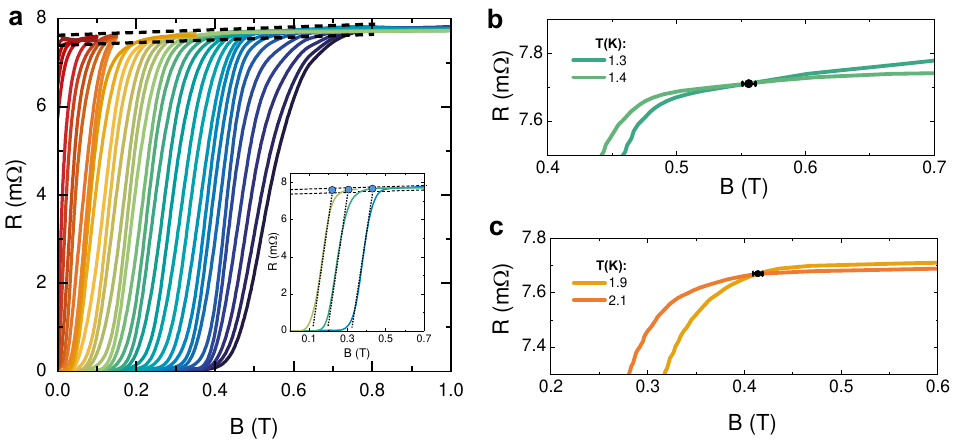}
	\end{center}\vspace{-15pt}
	\caption{
    \textbf{(a)} $\rho(B)$ isotherms obtained by matrix inversion of $\rho(T)$ isofield curves. Temperature ranges from 0.3\,K (black line) to\,4 K (dark red line) with increments of 0.1\,K, such that each isotherm corresponds to a specific temperature step.
    Dashed lines define the boundaries within which the material is considered to be in the normal state, as outlined in the main text.
    Representative examples of the determination of $B\ped{c}\apex{mf}$ using the intersection method are shown in the inset, where the curves correspond to temperatures of 1.2\,K (light blue line), 1.8\,K (green line), and 2.3\,K (light green line).
    The intersections with the normal region are marked by blue-filled points.
    \textbf{(b-c)} Representative examples of the determination of $B\ped{on}$ as the crossing point of two consecutive isotherms. }
	\label{fig:detailsBT}
\end{figure*}

\noindent To construct the $T-B$ phase diagram of H$_x$TiSe$_2$ from the magnetotransport measurements, we followed two different approaches.
In the first approach we considered the $\rho(T)$ isofield curves which were directly measured experimentally, and analyzed them as discussed in the Main Text.
In the second approach we estimated equivalent quantities from the $\rho(B)$ isotherms, which we determined from the isofield experimental data as follows.
Fist, we interpolated the measured $\rho(T)$ isofield curves onto a discrete temperature scale ranging from 0.3\,K to 4\,K, with a step size of 0.1\,K.
This interpolation allowed us to construct a matrix in which each entry corresponds to a resistivity value for a specific pair of magnetic field and temperature.
We then extracted the $\rho(B)$ isotherms, displayed in Supplementary Fig.\,\ref{fig:detailsBT}a, by slicing the matrix at fixed temperature values.

Next, we estimated the superconducting onset in the $T-B$ phase diagram by determining $B\ped{on}(T)$ from the $\rho(B)$ isotherms. 
Following previous studies\,\cite{Lewellyn2019PRB, Wang2023PRB}, we identified the crossing points between two adjacent curves and assigned the value of $B$ at the intersection as $B\ped{on}$.
The corresponding temperature was calculated as the average of the temperatures of the two isotherms, with the uncertainty defined as the semidispersion of these values.
The uncertainty in $B\ped{on}$ was instead estimated by considering the intrinsic noise level at the crossing point of the two curves involved.
This procedure was applied systematically for all temperatures (two examples are shown in Supplementary Fig.\,\ref{fig:detailsBT}b,c), resulting in the $B\ped{on}(T)$ points shown in Fig.\,2 and which exhibit a trend that closely matches that of the $T\ped{on}(B)$ points obtained from the isofield curves as discussed in the Main Text. \\
\indent Finally, to estimate the mean-field transition, we started by analyzing the measured $\rho(T)$ isofield curves.
We first identified the TAFF regime and then performed a linear fit in the region where $\rho(T)$ showed the highest linearity, as shown in Fig.\,1e.
This fitted line was extrapolated to intersect the curve measured at $B=1$\,T, where the material reaches the normal state.
Following the methodology of previous studies\,\cite{Berghuis1993PRB}, the temperature at this intersection was taken as $T\ped{c}\apex{mf}$ for the corresponding magnetic field of the analyzed curve, enabling us to obtain the $T\ped{c}\apex{mf}(B)$ data.
A similar method was applied to the $\rho(B)$ isotherms obtained by matrix inversion to determine the $B\ped{c}\apex{mf}(T)$ points. However, unlike the $\rho(T)$ analysis, a unique normal isotherm was not available as not all inverted isotherms span the entire magnetic field range, and the normal state displays a slight temperature dependence (see Fig.\,1c and 1e). To address this issue and ensure consistency, we treated the normal state as the envelope of all isotherms above $B\ped{on}$. The boundaries of this envelope are marked in Supplementary Fig.\,\ref{fig:detailsBT}a by the dashed black lines. Thus, the intersections were determined as the midpoint of the extrapolated curves and the envelope, with the uncertainty defined as the semi-difference of the values where the extrapolated curves intersect the dashed lines.
The two resulting mean-field data sets are displayed in Fig.\,2 and exhibit excellent agreement with each other.
We further note that this phenomenological procedure\,\cite{Berghuis1993PRB} provides results comparable to the more commonly used Ullah–Dorsey scaling theory\,\cite{Ullah1990PRL, Wang2023PRB}, while avoiding some of its limitations, such as failure below $T_c/2$ and the need for specific fluctuation theories or prior knowledge of the dimensionality of the system.

\clearpage

\section*{Supplementary Note 2: Fits of the PCARS spectra with a single anisotropic gap}

To highlight the difference in the spectral features associated with the opening of either two separate superconducting gaps, or a single anisotropic gap, on the Fermi surface, we focused on the PCAR spectra of $c$-axis contacts, since in these the shoulder located between 0.6 and 0.7\,meV is most evident.
For the anisotropic gap, we selected a standard angular dependency of the form:
\begin{equation}
    \Delta(\theta) = \Delta\ped{is} + \Delta\ped{an} \cos^4(2\theta)
\end{equation}
that smoothly evolves between the minimum value $\Delta\ped{is}$ and the maximum value $\Delta\ped{an}$ as a function of the angle $\theta$.
\begin{figure*}[h]
    \begin{center}
		\includegraphics[keepaspectratio, width=0.8\textwidth]{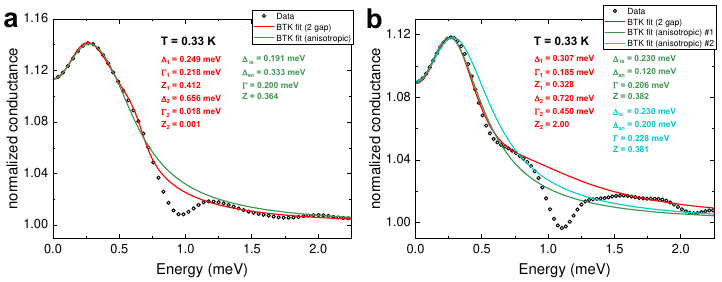}
	\end{center}\vspace{-10pt}
	\caption{
    Comparison between the fits to the experimental PCAR spectra (hollow black circles) with BTK models using two isotropic gaps (solid red lines) and one anisotropic gap (solid green and cyan lines).
    The PCAR spectra in panels \textbf{a} and \textbf{b} are those reported in Fig.\,3d and 3c respectively.
    The parameters of the BTK models are indicated in the labels.
    }
	\label{fig:anisotropic_gap}
\end{figure*}

As shown in Supplementary Fig.\,\ref{fig:anisotropic_gap}a for a first $c$-axis contact (that reported in Fig.\,3d of the Main Text), the two-gap model (solid red line) is able to reproduce well the spectrum at any energy below $\approx0.75$\,meV (above which critical current effects set in).
In particular, it captures well the zero-bias minimum, the maximum associated with the small gap at $\approx0.25$\,meV, and the shoulder at $\approx 0.65$\,meV.
Conversely, the anisotropic-gap model (solid green line) is able to reproduce well the experimental spectrum only up to energies $\approx 0.5$\,meV, that is it can only capture the zero-bias minimum and the maximum associated to the small gap.
At higher energies, it underestimates the experimental spectrum in the region of the large-gap shoulder and overestimates it in the energy range between that and the critical current "dip".

An even clearer example is shown in Supplementary Fig.\,\ref{fig:anisotropic_gap}b for a second $c$-axis contact (that reported in Fig.\,3c of the Main Text), where again the two-gap model (solid red line) is able to capture all the experimental spectral features up to the energies associated with the critical current effects -- and in particular the very broad shoulder around $\approx$0.75\,meV, which the anisotropic-gap model completely fails in reproducing within a least-squares best-fit procedure (solid green line).
If the least-squares constraint is lifted, different sets of parameters exist that allow the anisotropic-gap model to improve its agreement with the experimental curve in the shoulder region (solid cyan line); however, this improved agreement at higher energies comes at the expense of a reduced quality of the fit at zero energy and a large overestimation of the differential conductance in the energy range between the small-gap maximum and the large-gap shoulder.

Overall, these results clearly show that a standard model for an anisotropic gap fails in reproducing the fine details of our experimental PCAR spectra, which are instead very well captured by two separate isotropic gaps.

\clearpage

\section*{Supplementary Note 3: Additional PCARS spectra and fits including critical current effects}

\begin{figure*}[h]
    \begin{center}
		\includegraphics[keepaspectratio, width=\textwidth]{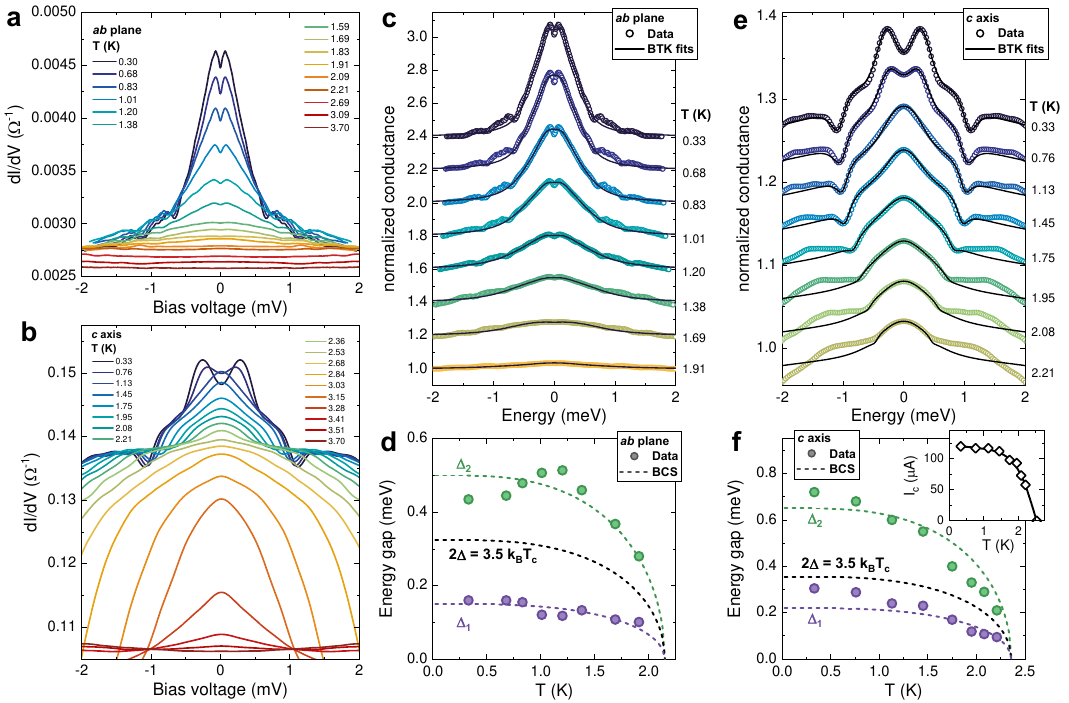}
	\end{center}\vspace{-10pt}
	\caption{  
            Additional examples of PCAR spectra.
            \textbf{a-b}, Symmetrized differential conductance curves ($dI/dV$ vs. $V$) measured in one $ab$-plane (\textbf{a}) and one $c$-axis (\textbf{c}) contact at different temperatures $T$.
            \textbf{c}, The curves of panel \textbf{a} after normalization (symbols) and the best-fitting curves (line) according to the two-band BTK model.
            \textbf{d}, $T$ dependence of $\Delta_1$ (purple circles) and $\Delta_2$ (green circles) extracted from the fits in panel \textbf{g}. Dashed lines are functions of the form $\Delta_i(T) = \Delta_i(0)\tanh \left(1.74 \sqrt{ T\ped{c}\apex{A}/T-1}\right)$.
            \textbf{e,f}, Same as \textbf{c,d} for the curves of panel \textbf{b}. The fits here have been performed with the extended BTK models which includes critical current effects.
            The inset to \textbf{f} shows the resulting $T$ dependence of the critical current $I\ped{c}$ in the contact (symbols).
            Solid line is a guide to the eye.
    }
	\label{fig:PCARS_and_dips}
\end{figure*}

Supplementary Figs.\,\ref{fig:PCARS_and_dips}a and \ref{fig:PCARS_and_dips}b report two examples of the $T$ dependencies of PCAR spectra in $ab$-plane and $c$-axis contacts, respectively, in addition to those provided in Fig.\,3g and 3h of the Main Text.
The low-temperature, high-voltage resistance of the contacts were approximately 350 $\Omega$ in the $ab$-plane case, and 7.4 $\Omega$ in the $c$-axis one. 

The spectra of the $ab$-plane contact (Supplementary Fig.\,\ref{fig:PCARS_and_dips}a) show a zero-bias dip and sharp peaks at $\pm 0.065$ mV that are the signature of a small nodeless gap; their overall shape is however completely incompatible with a single gap, indicating that at least another, larger gap is present.
This second gap is responsible for the width of the Andreev structures and for the  weak shoulder at about $\pm 0.4$\,mV.
The spectra do not present anomalous features and their temperature evolution is as expected for a ballistic contact, with just a small downward shift of the curves in correspondence with the resistive transition (which is due to the onset of the spreading resistance).
The fit of these curves, once normalized (and vertically offset by 0.2 for clarity) is shown in Supplementary Fig.\,\ref{fig:PCARS_and_dips}c.
The fit works fairly well at all temperatures even though it is unable to reproduce the persistence of the two clear conductance peaks up to 1.2\,K, because according to the model these maxima should be smeared out already at 0.68\,K.
The temperature dependence of the gap amplitudes is reported in Supplementary Fig.\,\ref{fig:PCARS_and_dips}d and shows that the small gap $\Delta_1$ approximately follows a BCS-like trend, while the large gap $\Delta_2$ displays a broad maximum slightly above 1\,K and then decreases in a BCS-like fashion.
Both the gaps close at the same temperature, equal to 2.15\,K.

The spectra of the $c$-axis contact clearly display the conductance maxima at about $\pm 0.3$ mV and clear shoulders around $\pm 0.6$ mV, which can be recognized as the hallmarks of the two gaps.
Moreover, sharp conductance dips are clearly visible at $\pm 1.1$ mV that move to lower voltages on increasing temperature, and a much larger downward shift of the curves occurs in correspondence to the resistive transition.
In this case, a proper extraction of the gap amplitudes requires taking into account the dips in the fitting function. 

The conductance dips arise when the radius $a$ of the point contact is not sufficiently small to ensure pure ballistic conduction ($a \ll \ell$, $\ell$ being the electron mean free path) but is neither so large to make the contact be completely ohmic ($a > \ell$).
In this intermediate condition, the resistance of a heterocontact between two metals (1 and 2) can be expressed by Wexler's formula:

\begin{equation}
R=R_s + \gamma\left (\frac{\ell}{a}\right )R_M = \frac{2h}{e^2 a^2 k_{F,min}^2 \tau}+
\gamma\left (\frac{\ell}{a}\right )\frac{\rho_1+\rho_2}{4a}\label{eq:Wexler}
\end{equation}
as the sum of two terms: the so-called Sharvin resistance, typical of ballistic contacts, which does not depend on temperature and is proportional to $a^{-2}$; and the so-called Maxwell term which is typical of Ohmic (dissipative) contacts and is proportional to $a^{-1}$.
The function $\gamma$ being always of the order of unity, the relative weight of the two contributions is governed by $a$: the smaller is $a$, the smaller is the Maxwell contribution.
The Maxwell term is proportional to the resistivity of the banks of the junction and, as long as the superconductor (bank 2) is in the zero-resistance state ($\rho_2=0$) this term can usually be neglected.
However, if the contact resistance is small, the current required to have $eV \geq \Delta$ is large and its density in the contact may easily exceed the critical value beyond which the zero-resistance state is destroyed.
In these conditions, a portion of the superconductor close to the point contact enters the resistive state, the Mawell term starts playing a role and the conductance suddenly decreases, giving rise to the dips. 
The experimental $V-I$ characteristic in these conditions is 

\begin{equation}
    V\ped{exp}(I) = V\ped{c}(I) + \int_0^I R_M(I') dI' \label{eq:Vcorr}
\end{equation}
where $V_c(I)$ is the voltage  drop at the contact alone, and $R_M$ is the current-dependent Maxwell contribution to the overall resistance, modelled by a proper function\,\cite{Daghero2023LTP} that is zero at zero current, displays a fast increase and a peak at the critical current $I_c$, and then a saturation to a non-zero value for $I>I_c$.
The total differential resistance turns out to be the sum of the resistance of the contact itself (with the structures associated to Andreev reflection) and of the Maxwell resistance.
Once inverted, this gives the differential conductance that, with a proper choice of the parameters, can be fitted to the experimental data.

Supplementary Fig.\,\ref{fig:PCARS_and_dips}e shows the experimental normalized curves (vertically shifted by 0.04 for clarity) with the relevant best-fitting curves.
At low temperatures, the agreement is extraordinarily good at any voltage; on increasing temperature, the fit still works well but only at low bias.
This is due to an additional effect (i.e. the progressive increase of the \emph{bulk} resistivity in the TAFF region) which is not included in the model and that makes the high-energy tails of the curves deviate from the model. 
The fitting procedure allows extracting the gap amplitudes, shown in Supplementary Fig.\,\ref{fig:PCARS_and_dips}f.
In this case, both the small and the large gap display a low-temperature linear behaviour rather than a saturation and are poorly represented by a BCS-like curve.
As in all other cases, the gaps close at the same temperature, here equal to 2.36 K.
The inset shows the temperature dependence of the critical current extracted from the fit of the experimental spectra, which turns out to be very similar to that reported in thin films of multi-gap superconducting TMD NbSe$_2$\,\cite{Ye2023Coatings}.

\clearpage

\section*{Supplementary Note 4: Temperature dependence of $\Gamma$ and $Z$ parameters}

\begin{figure*}[h]
    \begin{center}
		\includegraphics[keepaspectratio, width=0.8\textwidth]{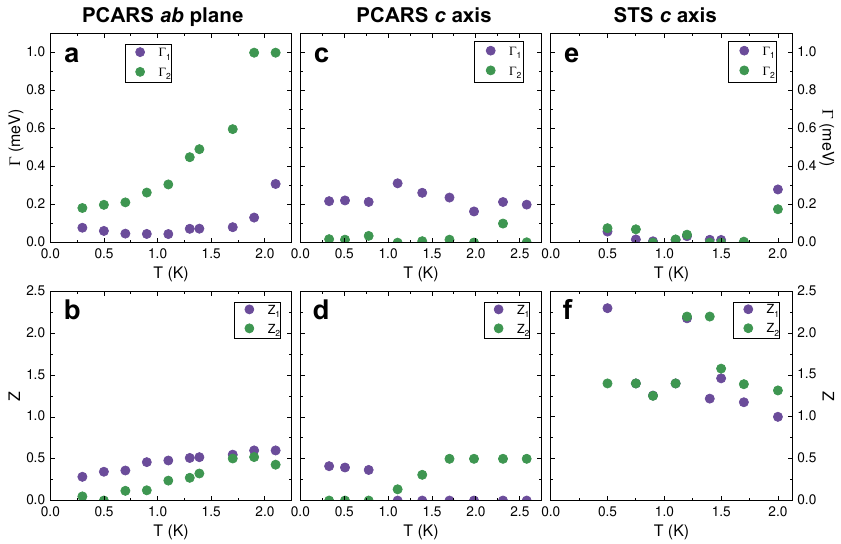}
	\end{center}\vspace{-10pt}
	\caption{Temperature dependencies of the broadening parameters $\Gamma$ and barrier transparencies $Z$ obtained from the fits of the experimental PCARS and STS curves with the two-band BTK models as discussed in the Main Text.
    Data in panels \textbf{a} and \textbf{b} refer to the fits shown in Fig.\,3g.
    Data in panels \textbf{c} and \textbf{d} refer to the fits shown in Fig.\,3i. 
    Data in panels \textbf{e} and \textbf{f} refer to the fits shown in Fig.\,4c.    
    }
	\label{fig:Gamma_Zeta}
\end{figure*}

Supplementary Figure\,\ref{fig:Gamma_Zeta} reports the temperature dependence of the broadening parameters $\Gamma_i$ (where $i=1.2$ is the gap index) and of the parameters $Z_i$ (that are proportional to the height of the potential barrier at the interface between the normal metal and the superconductor) for the $ab$-plane and  $c$-axis point contacts shown in the Main Text, as well as for the STS spectra. As clarified in the caption, panels a and b refer to the curves displayed in Fig.\,3g of the main text; panels c and d to the curves shown in Fig.\,3i of the main text; and panels e and f to the curves of Fig.\,4c.
In ideal contacts, the barrier parameters should not depend on temperature and should thus remain almost constant on increasing $T$, while the broadening parameters are usually allowed to increase with $T$. 
As for the point-contact spectra (panels a,b and c,d) the deviations from the ideal behavior can be mostly ascribed to the presence of dips that, moving towards lower bias voltages on increasing temperature (as a result of the decrease in the critical current) progressively distort the shape of the conductance curves.
In the case of the STS spectra, the variability of the $Z$ parameters likely arises from small modifications of the tunneling barrier as the sample is warmed up, which are greatly magnified by the low sensitivity of the shape of the BTK curves to variations in $Z_i$ values when $Z_i>1$. 

\clearpage

\section*{Supplementary Note 5: Fits of the STS spectra with a single-band Dynes model}

In tunneling spectroscopy, the differential conductance of single-gap superconducting materials are usually described by the equation
\begin{equation}
    \frac{dI(V)}{dV}=N\int_{-\infty}^{+\infty}\mathrm{Re}\left(\frac{E}{\sqrt{E^2-\Delta^2}}\right)\left[\frac{\partial f(E+eV)}{\partial eV}\right]dE
    \label{eq:Dynes}
\end{equation}
where $N$ is the constant density of states of the metallic tip, $\Delta$ is the superconducting gap, $f(E)$ is the Fermi distribution, and $E=E'-i\Gamma$ is the complex energy introduced by Dynes\,\cite{Dynes1978} which includes the spectral smearing parameter $\Gamma$ (analogous to that of the BTK model).
\begin{figure*}[h]
    \begin{center}
		\includegraphics[keepaspectratio, width=0.8\textwidth]{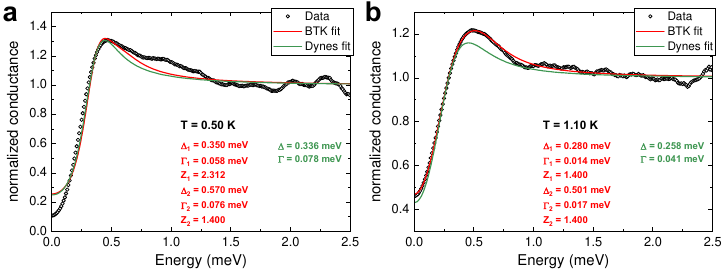}
	\end{center}\vspace{-10pt}
	\caption{
     Comparison between the fits to the experimental STS spectra (hollow black circles) with the two-gap BTK model (solid red lines) and the one-gap Dynes model (solid green lines; Eq.\,\eqref{eq:Dynes}).
    The tunneling spectra in panels \textbf{a} and \textbf{b} are those reported in Fig.\,4c and measured at $T=0.5$ and 1.1\,K respectively.
    The parameters of the BTK and Dynes models are indicated in the labels.
    }
	\label{fig:STS_single_gap}
\end{figure*}

In Supplementary Fig.\,\ref{fig:STS_single_gap} we show that Eq.\,\eqref{eq:Dynes} does not allow to capture the main features of our experimental tunneling curves.
If we consider the spectrum obtained at the lowest measured $T=0.5$\,K (panel a), we can see that the Dynes model (solid green line) is unable to describe the depth of the zero-bias minimum, as well as both the position of the maximum and the width of the coherence peak between $\approx 0.4$ and 0.5\,meV.
Conversely, the two-gap BTK model allows reproducing much more accurately the shape and position of the coherence peak, although it does not improve the agreement with the depth of the zero-bias minimum -- likely pointing to the existence of a broad distribution of (small) gap amplitudes as discussed in the Main Text.

The difference between the Dynes model and the two-gap BTK model in describing the experimental tunneling spectra becomes much more apparent at higher temperatures, such as the $T=1.1$\,K case displayed in panel b.
Here, the two-gap BTK model exhibits an excellent agreement with the experimental spectrum in the entire energy range.
Conversely, the Dynes model overestimates the depth of the zero-bias minimum and completely fails in accounting for the heigth, the position, and the width of the coherence peak. 

This comparison again confirms that the two-gap BTK model is the one that allows the best description of the experimental spectra in both the STS and PCARS measurements.

\clearpage

\section*{Supplementary Note 6: Two-band BCS model for $\Delta_{1,2}(T)$}
\noindent 
To assess whether the experimental temperature dependence of the gaps $\Delta_i$ can be accurately described by a two-gap model -- in particular, the quasi-linear $T$ dependence of the small gap $\Delta_1$ -- we first calculated its theoretically expected behavior by solving the self-consistent equations for a two-gap superconductor in the weak-coupling limit.

The original formulation was reported in the seminal work by Suhl, Matthias and Walker (SMW model)\,\cite{Suhl1959PRL}: 
\begin{equation}
\begin{aligned} 
&\Delta_1\left[1-V_{11}N_1(0)F(\Delta_1)\right] = \Delta_2V_{12}N_2(0)F(\Delta_2)\\
&\Delta_2\left[1-V_{22}N_2(0)F(\Delta_2)\right] = \Delta_1V_{12}N_1(0)F(\Delta_1)\\
%
\end{aligned}
\label{Suhl_eq_original}
\end{equation}
where $N_i(0)$ are the density of states at the Fermi level in band $i$, $V_{ii}$ are the intra-band interaction energies, $V_{12}=V_{21}$ is the inter-band interaction energy, and
\begin{equation}
    F(\Delta_{1,2})= \int^{\hbar\omega}_{0} d\varepsilon \tanh{\left(\frac{(\varepsilon^2+ \Delta_{1,2}^2)^{1/2}}{2\,k_B\,T}\right)}/\left(\varepsilon^2+ \Delta_{1,2}\right)^2
\end{equation}
where $T$ is the temperature.
Eqs.\,\eqref{Suhl_eq_original} can be rewritten in terms of the dimensionless BCS coupling constants\,\cite{Gurevich2003PRB} by defining $\lambda_{ij}=V_{ij}N_j$:
\begin{equation}
\begin{aligned} 
&\Delta_1= \Delta_1 F(\Delta_1)\,\lambda_{11}+ \Delta_2 F(\Delta_2)\, \lambda_{12} \\
&\Delta_2= \Delta_2 F(\Delta_2)\, \lambda_{22}+\Delta_1 F(\Delta_1)\, \lambda_{21} 
\end{aligned}
\label{Suhl_eq_lambda}
\end{equation}
where $\lambda_{11}$ and $\lambda_{22}$ are the intra-band BCS couplings, $\lambda_{21}$ and $\lambda_{12}$ are the inter-band BCS couplings.  These two equations, together with the expression for the critical temperature in the two-band BCS model\,\cite{Suhl1959PRL, Gurevich2003PRB}: 
\begin{equation}
k\ped{B}T\ped{c}= 1.14 \hbar\omega\ped{D} \exp\left[-\frac{\sqrt{4\lambda_{12}\lambda_{21}+(\lambda_{11}-\lambda_{22})^2}-(\lambda_{11}+\lambda_{22})}{2(\lambda_{12}\lambda_{21}-\lambda_{11}\lambda_{22})}\right]  
\label{eq:GurevichTc}
\end{equation}
and the fact that
\begin{equation}
    \lambda_{12}=\lambda_{21} \nu_{21} 
\end{equation}
where $\nu_{21}=N_2(0)/N_1(0)$ is the DOS ratio, provide a system of four equations in the unknown quantities $\lambda_{11}$, $\lambda_{22}$, $\lambda_{21}$ and $\lambda_{12}$ that can be solved by fixing $\nu_{21}$ and the upper integration limit $\hbar \omega$ in $F(\Delta_{1,2})$.
The latter corresponds to the maximum available phonon energy and can be estimated by assuming that the Debye temperature is approximately 203\,K\,\cite{Baranov2004CondMat}.

Operatively, we first solved this system at the lowest measured temperature (0.3 K) to extract the BCS couplings, using as input the critical temperature, the average gap values at 0.3 K, and a guess value of 0.35 for $\nu_{21}$.

The BCS couplings obtained in this way were then applied to iteratively solve Eq.\,\eqref{Suhl_eq_lambda} at progressively higher temperatures.  At each temperature, we started from the gap amplitudes determined at the previous temperature step, and recalculate them recursively by using Eqs.\,\eqref{Suhl_eq_rec}:
\begin{equation}
\begin{aligned} 
&\Delta_1^{i+1}= \Delta_1^{i} F(\Delta_1^{i})\,\lambda_{11}+ \Delta_2^{i} F(\Delta_2^{i})\, \lambda_{21}\, \nu_{21} \\
&\Delta_2^{i+1}= \Delta_2^{i} F(\Delta_2^{i})\, \lambda_{22}+\Delta_1^{i} F(\Delta_1^{i})\, \lambda_{21}
\label{Suhl_eq_rec}
\end{aligned}
\end{equation}
where $F(\Delta_{1,2})$ is calculated at the new temperature. Convergence of the calculation was achieved when the
difference between two successive iterations was less than a threshold of $10^{-12}$ meV.

The final result is the temperature dependence of the gaps, determined for a specific value of the gap ratio $\nu_{21}$. As expected, the shape of the $\Delta_1 (T)$ curve (in particular, close to the critical temperature) strongly depends on the initial choice of $\nu_{21}$. This parameter was then changed until we were able to reproduce the experimentally observed temperature dependence of the gaps, which was achieved for $\nu_{21}=0.48$. The corresponding values of the BCS couplings are reported in Table\,1 of the Main Text.

From these, we also calculated the effective total coupling constant of the system, $\lambda\ped{eff}$, by equating the expression for the critical temperature in the two-band BCS model, i.e. equation \ref{eq:GurevichTc}, 
%
%
to McMillan's formula for the critical temperature\,\cite{Dynes1972}:
\begin{equation}
k\ped{B}T\ped{c}= 1.14 \hbar\omega\ped{D} \exp\left(-\frac{1+\lambda\ped{eff}}{\lambda\ped{eff}-\mu^* }\right)    
\label{eq:McMillan}
\end{equation}
in which we set the Coulomb pseudopotential to zero (i.e. $\mu^* = 0$) for consistency with the two-band BCS theory, where its contribution is neglected\,\cite{Suhl1959PRL, Gurevich2003PRB}.
Following this procedure, we obtained the value of $\lambda\ped{eff} = 0.34$, as mentioned in the Main Text, which must be considered a lower bound to the actual coupling constant.

\clearpage

\section*{Supplementary Note 7: Two-gap Eliashberg model for $\Delta_{1,2}(T)$}

\noindent After establishing that the temperature dependence of the gaps in H$_x$TiSe$_2$ can be accurately described using a simple two-band BCS theory, we further investigated whether a comparable level of agreement could be achieved within the more comprehensive Eliashberg framework.
In the absence of magnetic impurities, the Eliashberg prediction for the temperature dependence of the gaps in a two-band system can be determined by solving the following set of four coupled equations for the gap functions $\Delta_i(i\omega_n)$ and the normalization functions $Z_i(i\omega_n)$ \cite{Carbotte1990RMP, Ummarino2009PRB, Ghigo2017PRB, Ummarino2004PC, Ummarino1997PRB, Torsello2019PRB}:
\begin{gather} 
Z_{i}(i\omega_{n})\Delta_{i}(i\omega_{n})=\pi
T\sum_{m,j}\big[\Lambda_{ij}(i\omega_{n},i\omega_{m})-\mu^{*}_{ij}(\omega_{c})\big]\,\,\Theta(\omega_{c}-|\omega_{m}|)\,\,N^{\Delta}_{j}(i\omega_{m})
+\sum_{j}\Gamma^{N}_{ij}\,N^{\Delta}_{j}(i\omega_{n}) \nonumber\\
\label{eq_ELiashbergB0}
\omega_{n}Z_{i}(i\omega_{n})=\omega_{n}+ \pi T\sum_{m,j}\Lambda_{ij}(i\omega_{n},i\omega_{m})\,N^{Z}_{j}(i\omega_{m})+\sum_{j} \,\Gamma^{N}_{ij}\,N^{Z}_{j}(i\omega_{n})\\ 
\Lambda_{ij}(i\omega_{n},i\omega_{m})=2
\int_{0}^{+\infty} \frac{\alpha^{2}_{ij}(\Omega)F(\Omega)}{(\omega_{n}-\omega_{m})^{2}+\Omega^{2}} \,\Omega\,d\Omega \nonumber
\end{gather}
where \textit{i} denotes the band index, $\omega_n$ are the Matsubara frequencies, $\omega_c$ is a cutoff energy, $\Theta$ is the Heaviside step function, and $N^{\Delta/Z}_{j}(i\omega_{n})$ are defined as $ N^{\Delta}_{j}(i\omega_{n})=\Delta_{j}(i\omega_{n})/
{\sqrt{\omega^{2}_{n}+\Delta^{2}_{j}(i\omega_{n})}}$ and $N^{Z}_{j}(i\omega_{n})=\omega_{n}/{\sqrt{\omega^{2}_{n}+\Delta^{2}_{j}(i\omega_{n})}}$, respectively.
Solving Eqs.\,\eqref{eq_ELiashbergB0} requires a significant number of input parameters: four elements of the Coulomb pseudopotential matrix $\mu_{ij}(\omega_c)$, four nonmagnetic scattering rates $\Gamma_{ij}^N$, and four electron-phonon spectral functions $\alpha^2_{ij}(\Omega)F(\Omega)$.
From these, the elements of the electron-phonon coupling matrix would then be calculated as $\lambda_{ij}=2\int_0^{+\infty}\alpha^{2}_{ij}(\Omega)F(\Omega)/\Omega\,d\Omega$.
Nonetheless, some of these parameters can be derived from experimental measurements, while others can be determined through reasonable approximations.

First, we assumed that the shape of the electron-phonon functions is the same for both intra-band and inter-band interactions, with their respective weights captured by the coupling constants $\lambda_{ij}$.
This allows us to express them as $\alpha_{ij}^2(\Omega)F(\Omega) = \lambda_{ij} \alpha^2F(\Omega)$, and to set $\alpha^2F(\Omega)$ to be that determined in H$_x$TiSe\ped{2} via ab initio calculations\,\cite{Piatti2023CommunPhys}.
This approach eliminates the need for four separate spectral functions, reducing the problem to the determination of only three $\lambda_{ij}$ and of the DOS ratio $\nu_{21}$, since also for the Eliashberg couplings it holds that $\lambda_{12}=\lambda_{21}\nu_{21}$.

Second, to further reduce the number of free parameters involved, we set $\mu_{ij}^{*}$ to zero -- as inherently assumed by the BCS two-band model\,\cite{Suhl1959PRL, Gurevich2003PRB} -- thereby obtaining a lower bound for the electron-phonon coupling constants.

Finally, the nonmagnetic scattering rates $\Gamma_{ij}^N$ were also set to zero.
This decision was based on the observation that inter-band scattering leads to a strong isotropization of the two gaps and of their temperature dependencies, at variance with the well-separated values and trends observed experimentally.

Using these approximations, we solved Eqs.\,\eqref{eq_ELiashbergB0} with a cutoff energy $\omega_c=340$ meV, a maximum quasiparticle energy $\omega_{max}=350$ meV and a DOS ratio $\nu_{21}=0.48$. 
Again, the set of Eliashberg couplings $\lambda_{ij}$ that allow to best reproduce the experimental data are reported in Table\,1 of the Main Text.
We then computed the total effective coupling constant, which in the two-band Eliashberg theory is simply
\begin{equation}
\lambda\ped{eff}= [\lambda_{11}+\lambda_{12}+\nu_{21}(\lambda_{22}+\lambda_{21})]/(1+\nu_{21})
\end{equation}
yielding a value $\lambda\ped{eff}=0.34$, in excellent agreement with the weak-coupling BCS model.

\clearpage

\section*{Supplementary Note 8: Two-gap BCS model for $B_{c2}(T)$}

\noindent The temperature dependence of the upper critical field has been first fitted to the weak-coupling BCS model for a two-gap superconductor in the dirty limit originally derived by Gurevich\,\cite{Gurevich2003PRB}.
This model neglects inter-band scattering and accounts for intra-band contributions through a diffusivity parameter $D_i$, specific to each band.
Within this framework, the relationship between $B\ped{c2}$ and temperature is given by the implicit equation:
\begin{equation}
a_0[\ln(t)+ U(b/t)][\ln(t) + U(\eta_{21} b/t)]+ a_1[\ln(t)+ U(b/t)]+ a_2[\ln(t) + U(\eta_{21} b/t)]= 0
\end{equation}
where $t=T/T\ped{c}$ is the reduced temperature, $b(T)= B\ped{c2}(T)/B\ped{c2}(0)$ the normalized upper critical field, and $\eta_{21}= D_2/D_1$ is the diffusivity ratio of the two bands. The function $U(x)$ is defined as $\Psi(0.5+x)-\Psi(x)$, where $\Psi(x)$ is the digamma function and the multiplicative coefficients $a_0$, $a_1$, and $a_2$ are related to the BCS coupling constants as follows: $a_0= 2\,(\lambda_{11}\lambda_{22}-\lambda_{12}\lambda_{21})/\lambda_{0}$, $a_1= 1+(\lambda_{11}-\lambda_{22})/\lambda_{0}$, $a_2=1-(\lambda_{11}-\lambda_{22})/\lambda_{0}$ where $\lambda_{0}^2= (\lambda_{11}-\lambda_{22})^2-4\,\lambda_{12}\lambda_{21}$.

$B\ped{c2}(0)$ and $\eta_{21}$ were treated as free parameters in the fitting procedure and their optimal values ($0.79$\,T and $12.36$ respectively) were obtained by performing a nonlinear regression of the experimental data, iteratively minimizing the least squares.

\clearpage

\section*{Supplementary Note 9: Two-gap Eliashberg model for $B_{c2}(T)$ }

\noindent As we did for the gaps, we also explored the temperature dependence of $B_{c2}$ within the more exhaustive Eliashberg framework. In this case, when a magnetic field is applied, Eqs.\,\eqref{eq_ELiashbergB0} take the following linearized form:
\begin{gather}
Z_{i}(i\omega_{n})\Delta_{i}(i\omega_{n})= \pi
T\sum_{m,j}\,\,\left\{\left[\Lambda_{ij}(i\omega_{n}-i\omega_{m})-\mu^{*}_{ij}(\omega_{c})\right]\, \Theta(\omega_{c}-|\omega_{m}|)+\delta_{n,m}\frac{\Gamma^{N}_{ij}}{\pi T}\right\}\,\,\chi_{j}(i\omega_{m})\,\,Z_{j}(i\omega_{m})\,\,\Delta_{j}(i\omega_{m})\nonumber\\
\label{eq:EliashbergB}
\chi_{j}(i\omega_{m})= \frac{2}{\sqrt{\beta_{j}}}\int^{+\infty}_{0}
\tan^{-1}\left[\frac{q\sqrt{\beta_{j}}}{|\omega_{m}Z_{j}(i\omega_{m})|+i\mu_{B}B_{c2}\mathrm{sign}(\omega_{m})}\right]\,e^{-q^{2}}\,dq\\
\omega_{n}Z_{i}(i\omega_{n}) = \omega_{n}+\pi
T\sum_{m,j}\,\,\left[\Lambda_{ij}(i\omega_{n}-i\omega_{m})+\delta_{n,m}\frac{\Gamma^{N}_{ij}}{\pi T}\right]\,\,\mathrm{sign}(\omega_{m})\nonumber 
\end{gather}
where $\beta_{i}=\pi B_{c2} v_{Fi}^{2}/(2\Phi_{0})$ and $v_{Fi}$ is the Fermi velocity in the $i$-th band. 

\begin{figure*}[b]
    \begin{center}
		\includegraphics[keepaspectratio, width=0.5\textwidth]{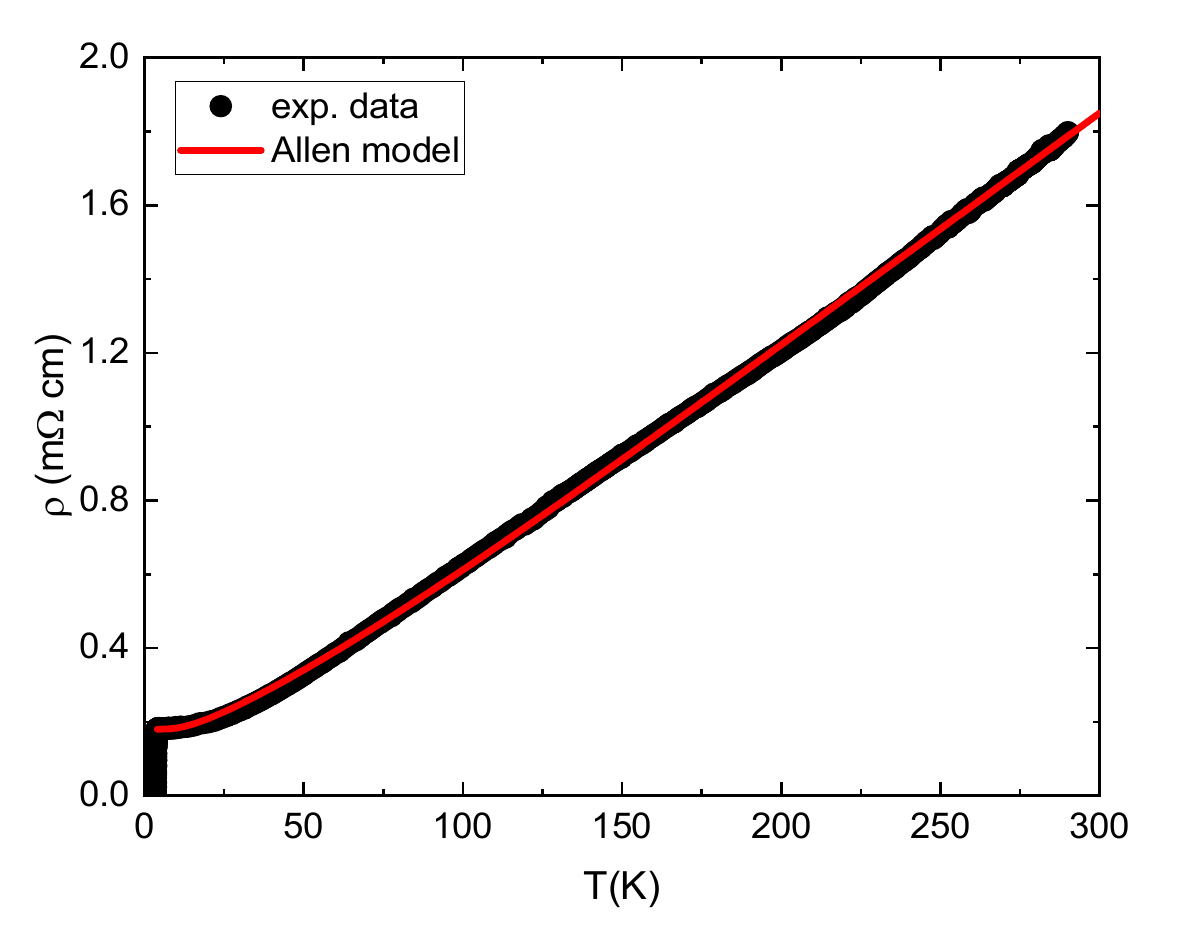}
	\end{center}\vspace{-15pt}
	\caption{Experimental temperature dependence of the resistivity in H$_x$TiSe$_2$ (black circles) along with its best-fit curve (solid red line) obtained from Allen theory\,\cite{Allen1978PRB}.}
	\label{fig:gammaGianni}
\end{figure*}

To minimize the number of free parameters in Eq.\,\eqref{eq:EliashbergB}, we estimated the nonmagnetic scattering rates by examining the temperature dependence of the resistivity in the normal state.
For a two-band system without magnetic impurities, this dependence is expected to follow the Allen theory\,\cite{Allen1978PRB}, which states:
\begin{equation}
\begin{aligned} 
	\frac{1}{\rho(T)} = & \frac{\varepsilon_0(\hbar\omega_{p})}{\hbar} \sum_{i=1,2}
	\frac{(\omega_{p,i}/\omega_{p})^2}{\gamma_i^N+W_i(T)}\\
   W_i(T) =4\pi k_BT\int_0^\infty 	&		\left[\frac{\hbar\Omega/2k_BT}{\sinh\big(\hbar\Omega/2k_BT\big)}\right]^2
			\frac{\alpha_{tr,i}^2(\Omega)F_{tr}(\Omega)}{\Omega} \,d\Omega
\end{aligned}
\end{equation}
where $\omega_{p,i}$ is the bare plasma frequency of the i-th band, $\omega_p$ is the total bare plasma frequency, and $\gamma_i$ is the sum of all intra-band and inter-band scattering rates due to nonmagnetic impurities.
For consistency with the BCS treatment of $B_{c2}(T)$, which assumes negligible inter-band contributions, inter-band scattering is specifically neglected.

The term $\alpha_{tr,i}^{2}F_{tr}(\Omega)= \sum_{j}\alpha_{tr,ij}^{2}F_{tr}(\Omega)$ refers to the sum of all transport spectral functions for the $i$-th band.
As previously assumed, the transport spectral functions share the same shape across both bands, irrespective of whether the interactions are intra-band or inter-band.
Their respective weights are incorporated through the coupling constants $\lambda_{tr,ij}$, leading to the relation $\alpha_{tr,ij}^2(\Omega)F_{tr}(\Omega) = \lambda_{tr,ij} \alpha_{tr}^2F_{tr}(\Omega)$.
These transport spectral functions are analogous to the standard Eliashberg $\alpha^2F(\Omega)$, with a key distinction as $\Omega \rightarrow 0$: $\alpha_{tr}^{2}F_{tr}(\Omega)$ behaves as $\Omega^4$, whereas $\alpha^2F(\Omega)$ behaves as $\Omega^2$.
The condition $\alpha_{tr}^{2}F_{tr}(\Omega) \propto \Omega^4$ should be imposed in the range $0 < \Omega < \Omega_D$, where $\Omega_D \approx \Omega\ped{log}/10$.
Consequently, we construct $\alpha^2_{tr}(\Omega)F_{tr}(\Omega)=b_{i}\Omega^{4}\vartheta(\Omega_{D}-\Omega)+c_{i}\alpha^2(\Omega)F(\Omega)\vartheta(\Omega-\Omega_{D})$, with the coefficients $b_i$ and $c_i$ determined by ensuring continuity at $\Omega_D$ and normalization.

At this stage, the free parameters in this model include the electron-phonon coupling constants $\lambda_{1,tr}$ and $\lambda_{2,tr}$, the impurities contents $\Gamma^{N}_1$ and $\Gamma^{N}_2$, and the plasma energies $\omega_{p,1}$ and $\omega_{p,2}$.
By fitting the experimental resistivity in the normal state (see Supplementary Fig.\,\ref{fig:gammaGianni}), we obtain the best-fit parameter values: $\omega_{p,1}=180$ meV, $\omega_{p,2}=517$ meV, $\Gamma^{N}_1=2.1$ meV, $\Gamma^{N}_2=9.0$ meV, $\lambda_{1,tr}=0.25$ and $\lambda_{2,tr}=0.45$.
The resulting total electron-phonon effective coupling constant in the normal state, calculated as $\lambda\ped{eff,tr}=(\lambda_{1}+\lambda_{2}\nu_{21})/(1+\nu_{21})$, is equal to 0.32.
Thus, $\lambda\ped{eff} \geq \lambda\ped{eff,tr}$, as expected for a standard electron-phonon superconductor.

Using these scattering rates, we solved Eqs.\,\eqref{eq:EliashbergB} by adjusting the two remaining free parameters, i.e. the bare Fermi velocities of the two bands. A convergent solution was found for $v_{F1} = 0.50\cdot10^5$\,m/s, and $v_{F2} = 3.35\cdot10^5$\,m/s, consistent with ab initio calculations\,\cite{Piatti2023CommunPhys}.
From these, the diffusivities of the two bands can be determined as $D_i\propto v^2\ped{F,\textit{i}}/\Gamma^N_i$ and yield a diffusivity ratio $\eta_{21} = 10.47$, in good agreement with the value determined with the Gurevich model.